\documentclass[fleqn,usenatbib]{mnras}

\usepackage{lipsum, babel}
\usepackage{newtxtext,newtxmath}
\usepackage[figuresright]{rotating}
\usepackage{amsmath}	

\usepackage{amssymb}	
\usepackage[export]{adjustbox}
\usepackage{multirow}
\usepackage[T1]{fontenc}
\usepackage{appendix}
\usepackage{enumitem}
\usepackage{graphicx,url}
\usepackage{color}
\usepackage[colorinlistoftodos]{todonotes}
\usepackage{xcolor}

\renewcommand{\arraystretch}{1.5}
\DeclareRobustCommand{\VAN}[3]{#2}
\let\VANthebibliography\thebibliography
\def\thebibliography{\DeclareRobustCommand{\VAN}[3]{##3}\VANthebibliography}
\usepackage{amsmath} 
\newcommand{\angstrom}{\text{\normalfont\AA}}

\def\xmm{{\it XMM-Newton}}

\def\athena{{\it ATHENA}}
\def\xrism{{\it XRISM}}
\def\chandra{{\it Chandra}}

\title[Winds in Mrk 1044]{Constraints on the ultra-fast outflows in the narrow-line Seyfert 1 galaxy Mrk 1044 from high-resolution time- and flux-resolved spectroscopy}
\author[Y. Xu et al.]{Yerong Xu$^{1,2}$,\thanks{E-mail: yerong.xu@inaf.it}
    Ciro Pinto$^{1}$, Daniele Rogantini$^{3}$, Stefano Bianchi$^{4}$, Matteo Guainazzi$^{5}$, Erin Kara$^{3}$,\newauthor Chichuan Jin$^{6,7}$, Giancarlo Cusumano$^{1}$,
\\
$^{1}$INAF - IASF Palermo, Via U. La Malfa 153, I-90146 Palermo, Italy\\
$^{2}$Universit\`a degli Studi di Palermo, Dipartimento di Fisica e Chimica, via Archirafi 36, I-90123 Palermo, Italy\\
$^{3}$MIT Kavli Institute for Astrophysics and Space Research, Massachusetts Institute of Technology, Cambridge, MA 02139, USA\\
$^{4}$Dipartimento di Matematica e Fisica, Università degli Studi Roma Tre, via della Vasca Navale 84, I-00146 Roma, Italy\\
$^{5}$ESA European Space Research and Technology Centre (ESTEC), Keplerlaan 1, 2201 AZ, Noordwĳk, The Netherlands\\
$^{6}$National Astronomical Observatories, Chinese Academy of Sciences, 20A Datun Road, Beijing 100101, China\\
$^{7}$School of Astronomy and Space Sciences, University of Chinese Academy of Sciences, 19A Yuquan Road, Beijing 100049, China
}

\date{Accepted 2023 May 19, Received 2023 May 19, in original form 2023 May March 22}

\pubyear{2023}

\begin{document}
\label{firstpage}
\pagerange{\pageref{firstpage}--\pageref{lastpage}}
\maketitle

\begin{abstract}
Ultra-fast outflows (UFOs) have been revealed in a large number of active galactic nuclei (AGN) and are regarded as promising candidates for AGN feedback on the host galaxy. The nature and launching mechanism of UFOs are not yet fully understood. Here we perform a time- and flux-resolved X-ray spectroscopy on four \xmm\ observations of a highly accreting narrow-line Seyfert 1 (NLS1) galaxy, Mrk 1044, to study the dependence of the outflow properties on the source luminosity. We find that the UFO in Mrk 1044 responds to the source variability quickly and its velocity increases with the X-ray flux, suggesting a high-density ($10^{9}\mbox{--}4.5\times10^{12}\,\mathrm{cm}^{-3}$) and radiatively driven outflow, launched from the region within a distance of $98\mbox{--}6600\, R_\mathrm{g}$ from the black hole. The kinetic energy of the UFO is conservatively estimated ($L_\mathrm{UFO}\sim4.4\%L_\mathrm{Edd}$), reaching the theoretical criterion to affect the evolution of the host galaxy. We also find emission lines, from a large-scale region, have a blueshift of $2700\mbox{--}4500$\,km/s in the spectra of Mrk 1044, which is rarely observed in AGN. By comparing with other sources, we propose a correlation between the blueshift of emission lines and the source accretion rate, which can be verified by a future sample study.

\end{abstract}




\section{Introduction}
It is well accepted that active galactic nuclei (AGN) are powered by the accretion of matter onto supermassive black holes (SMBHs) in the hearts of galaxies. The energetic output of AGN can impact the evolution of their host galaxies, an effect that is referred to AGN feedback \citep[e.g.][and references therein]{2012Fabian}. The enormous amount of energy and momentum, released in the form of matter and radiation, can expel or heat the surrounding interstellar medium (ISM). This may delay the gas cooling and further leads to the star formation (SF) quenching \citep{2012Zubovas}. In the early phases of feedback, AGN outflows can also trigger the star formation within the compressed gas \citep[e.g.][]{2017Maiolino}. Ultra-fast outflows (UFOs) with a wide solid angle are now considered one of the main mechanisms of AGN feedback for their mildly relativistic speeds ($\geq10000$\,km/s or $0.03c$) and powerful kinetic energy ($\geq0.05L_\mathrm{Edd}$). Such a huge kinetic output matches the theoretical predictions of effective AGN feedback models \citep[e.g.][]{2005DiMatteo,2010Hopkins}, offering an interpretation of the observed AGN-host galaxy relations \citep[e.g. $M_\mathrm{BH}-\sigma$,][and references therein]{2013Kormendy}.

UFOs are commonly detected by identified blueshifted Fe \textsc{xxv/xxvi} absorption lines above $7$\,keV in the X-ray band \citep[e.g.][]{2002Chartas,2006Cappi,2010Tombesi,2013Tombesi,2013Gofford,2022Matzeu}. The measured velocities of UFOs can reach up to $\sim0.3c$ \citep[e.g. APM 08279+5255 and PDS 456,][]{2002Chartas, 2003Reeves}, implying that they likely originate from the inner region of the accretion disk within several hundred gravitational radii from the black hole. Thanks to the high spectral resolution of the Reflection Grating Spectrometer \citep[RGS,][]{2001denHerder} onboard \xmm\ \citep{2001Jansen} and the High Energy Transmission Gratings \citep[HETG,][]{2005Canizares} onboard \chandra\ \citep{2002Weisskopf}, UFOs are also detectable in soft X-ray bands and distinguishable from slow, ionized outflows, the so-called warm absorbers.

Under the investigation of the past two decades, UFOs show variable signatures, i.e., variable velocities and transient features, based on multi-epoch deep observations \citep[e.g.][]{2012Dauser,2017Matzeu,2020Igo}. However, the exact nature of UFO variability and their launching mechanisms are not well understood. They could be driven either by the radiation pressure \citep[e.g.][]{2000Proga,2010Sim,2016Hagino} or by magneto-rotational forces \citep[MHD models, e.g.][]{2004Kato,2010Fukumura,2015Fukumura} or a combination of both. Variability might be key to determining UFO launching mechanisms. It has been found that the Fe K absorption features in the spectrum of IRAS 13244-3809 and 1H 0707-495 weaken with the increasing X-ray luminosity, implying an over-ionization of the gas \citep{2017Parker,2018Pinto,2021Xu}. The velocity of the UFO in PDS 456 and IRAS 13229-3809 increases with the source luminosity \citep{2017Matzeu,2018Pinto}. The above discoveries support that UFOs in high-accretion systems are mainly accelerated by the strong radiation field. Interestingly, \citet{2021Xu} found, instead, an anti-correlation between the UFO velocity and X-ray luminosity in 1H 0707-495, challenging our understanding of the UFO driving mechanism. It was explained by the supercritical flow expanding at high-accretion states, resulting in larger launching radii (i.e. at lower velocities) within the disk. Therefore, it is worth investigating the dependence of UFOs on the source luminosity and accretion rate in other sources to better understand the nature of UFOs.

Mrk 1044 is a nearby ($z=0.016$) and luminous \citep[$L_\mathrm{1\mu m\mbox{--}2keV}=1.4\times10^{44}\,\mathrm{erg/s}$,][]{2010Grupe} narrow-line Seyfert 1 AGN, hosting a central SMBH with a reverberation-mapped mass of $M_\mathrm{BH}=2.8\times10^6\,M_\odot$ \citep{2015Du} or a mass, determined through the FWHM(H$\beta$) and $L_{5100\angstrom}$, of $M_\mathrm{BH}=2.1\times10^6\,M_\odot$ \citep{2010Grupe}. Mrk 1044 shows a soft X-ray excess in the spectrum \citep{2007Dewangan}. It was interpreted by relativistic reflection from a high-density accretion disk in \citet{2018Mallick}, although in general a warm corona model also provides a statistically acceptable description of the soft excess below 2\,keV \citep[e.g.][]{2018Petrucci,2019Garc,2020Petrucci,2021XuSE}. In the \xmm/RGS spectrum, based on a series of narrow absorption lines, \citet{2021Krongold} found four distinctive UFOs, explained by a shocked outflow scenario. From the multi-wavelength observations, Mrk 1044 was reported to have multi-phase outflows in optical and UV bands as well, including two unresolved and one resolved ionized gas outflows traced by [O \textsc{iii}] in the optical band as well as two Ly-$\alpha$ absorbing components in the ultra-violet (UV) energy range \citep{2005Fields,2022Winkel}. 

In this paper, we will present the high-resolution spectroscopic analysis on four \xmm/RGS observations of Mrk 1044 (PI: C. Jin). In section \ref{sec:data}, we present the four \xmm\ observations and our data reduction process. Details on our analysis and results are shown in section \ref{sec:results}, where we expand the work by \citet{2021Krongold}, find an additional blueshifted photoionized emission component, and further study the relation between the properties of winds and the source luminosity. We discuss the results and provide our conclusions in section \ref{sec:discussion} and section \ref{sec:conclusion}, respectively.

\section{Data reduction and products}\label{sec:data}
Mrk 1044 has been observed with a large \xmm\ program (PI: C. Jin) for three orbits in 2018 and one orbit in 2019. The details of the analyzed observations in this work are listed in Tab.\ref{tab:obslog}. \xmm\ consists of the European Photon Imaging Camera (EPIC), including two EPIC-MOS CCDs \citep{2001Turner} and an EPIC-pn \citep{2001Struder}, RGS, and the Optical Monitor \citep[OM,][]{2001Jansen}. This work focuses on the RGS and we use the EPIC and OM data mainly to determine the shape of the broadband spectral energy distribution (SED), for which the MOS results are redundant as pn has a significantly higher effective area in the hard band.
\begin{table}
\centering
\caption{General overview of the analyzed observations on Mrk 1044
}
\begin{tabular}{lcccccccccc}
\hline
\hline
Obs. ID & Date & Instrument & Net exp. & Net count rate \\
&&&(ks)&(cts/s)\\
\hline
\multirow{2}{*}{0824080301} & \multirow{2}{*}{2018-08-03} & EPIC-pn & 95 & 32\\
&&RGS& 134 & 1.07\\
\multirow{2}{*}{0824080401} & \multirow{2}{*}{2018-08-05} & EPIC-pn & 97 & 24\\
&&RGS& 133 & 0.79\\
\multirow{2}{*}{0824080501} & \multirow{2}{*}{2018-08-07} & EPIC-pn & 93 & 25\\
&&RGS& 131 & 0.84\\
\multirow{2}{*}{0841820201} & \multirow{2}{*}{2019-08-03} & EPIC-pn & 90 & 20\\
&&RGS& 126 & 0.63\\
\hline
\end{tabular}
\label{tab:obslog}
\end{table}

\subsection{Data reduction}\label{subsec:reduction}
The data sets are processed with the \xmm\ Science Analysis System (SAS v20.0.0) and calibration files available by September 2022, following the standard SAS threads. We reduced EPIC-pn data using \textsc{epproc} package and produced calibrated photon event files. The filter of the background flare contamination is set at 0.5 counts/sec in 10-12 keV. We extracted the source spectra from a circular region of radius 30 arcsec, and the background spectra from a nearby source-free circular region with the same radius. No significant pile-up effect is found with the task \textsc{epatplot}. The EPIC-pn spectra are grouped to over-sample the instrumental resolution at least by a factor of 3 and each energy bin has a minimum of 25 counts to maximize the S/N. We employed the \textsc{rgsproc} package to process the RGS data with a filter of 0.3 counts/sec to exclude the background flares. The first-order RGS spectra are extracted from a cross-dispersion region of 1 arcmin width. The background spectra are selected from photons beyond 98\% of the source point-spread function. The RGS1 and RGS2 spectra are combined and grouped to over-sample the resolution at least by a factor of 3. During the observation, Mrk 1044 was also monitored by OM in the UVW1 (2910\AA) filter. We reduced OM data with \textsc{omichain} tool including all necessary calibration processes. The response file is retrieved from the ESA webpage\footnote{https://www.cosmos.esa.int/web/xmm-newton/om-response-files}. The UVW1 flux is less variable than the X-ray flux, i.e., almost stable in 2018 and drops by 13\% in 2019.

\begin{figure*}
	\includegraphics[width=\textwidth, trim={50 50 30 20}]{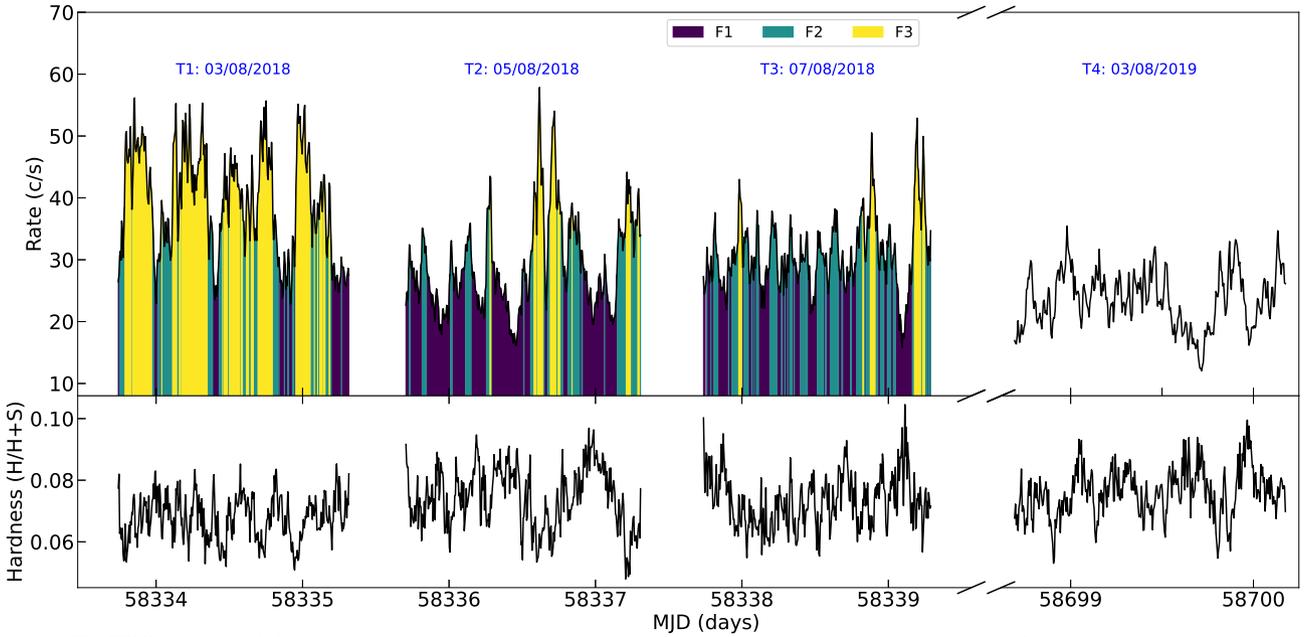}
    \caption{The EPIC-pn ($0.3\mbox{--}10$\,keV) light curve ({\it upper}) and corresponding hardness ratio ({\it lower}) of the observations of Mrk 1044, where the observation dates (T1-T4) are marked. The colors represent the different flux intervals (F1-F3) with comparable counts.}
    \label{fig:lc}
\end{figure*}

\subsection{Light curve}\label{subsec:lightcurve}
By using the task \textsc{epiclccorr}, we present the background-substracted and deadtime-corrected light curves extracted from the EPIC-pn ($0.3\mbox{--}10$\,keV) data in Fig.\ref{fig:lc}. It reveals that Mrk 1044 is bright and variable during the observations. The corresponding hardness ratio (HR=H/H+S, H: $2\mbox{--}10$\,keV; S: $0.3\mbox{--}2$\,keV), plotted in the bottom panel, shows a softer-when-brighter behavior. To investigate the variability of the UFO with the luminosity, we divide three consecutive observations in 2018 into three flux levels, marked by different colors. The reason why we exclude the 2019 observation is to ensure the causality between the variations of the UFO and the luminosity, i.e., we are studying the response of the same absorber to the source. The thresholds are set to make the number of counts of each level comparable. The good time interval (GTI) files for each level are generated with \textsc{tabgtigen} task. The flux-resolved EPIC-pn and RGS spectra at the same flux level are extracted and stacked following the steps described in Sec.\ref{subsec:reduction}. The observations in 2018 are also stacked into one single spectrum, named 2018. In this work, we will perform the flux-/time-resolved spectroscopy for a total of 8 spectra, where the time-resolved spectra are referred to as T1...T4 chronologically (e.g. T1 refers to Obs. 0824080301) and the flux-resolved spectra are referred to as F1, F2, F3 from the lowest to the highest state.

\section{Results}\label{sec:results}
\subsection{Continuum Modelling}\label{subsec:continuum}
We start the broadband X-ray spectroscopy from the stacked 2018 EPIC-pn and RGS spectra, due to their high statistics, in the XSPEC (v12.12.1) package \citep{1996Arnaud}. The instrumental differences are accounted for  by adopting a variable cross-calibration factor. In this paper, we use the $\chi^2$ statistics and estimate the uncertainties of all parameters at the default 90\% confidence level (i.e. $\Delta\chi^2=2.71$), but 1$\sigma$ ($\Delta\chi^2=1$) error bars are shown in plots. We consider the RGS spectra between $0.4\mbox{--}1.77$\,keV and the EPIC-pn spectra between $1.77\mbox{--}10$\,keV in our analysis, not only because of their consistency in the soft X-ray band, but also due to the influence of the lower resolution but higher count rate of EPIC-pn on the detection of atomic features. The luminosity calculations in this paper are based on the assumptions of $H_0=70\,\mathrm{km/s/Mpc}$, $\Omega_\Lambda=0.73$ and $\Omega_M=0.27$. 

The broadband X-ray model for Mrk 1044 was proposed by \citet{2018Mallick} based on the archival \xmm\ data in 2013. In this work, we adopt a similar model combination: \texttt{tbabs*zashift*(nthComp+relxilllpCp)}, to explain those spectral components. The model takes into account the galactic hydrogen absorption (\texttt{tbabs}) with the solar abundance calculated by \citet{2009Lodders}, the redshift of Mrk 1044 (\texttt{zashift}), the soft excess in form of a warm Comptonization component (\texttt{nthComp}), and the hot coronal continuum like a power-law plus a lamppost-geometry relativistic reflection \citep[\texttt{relxilllpCp}, RELXILL v1.4.3,][]{2014Garc}. The Galactic column density, $N_\mathrm{H}^\mathrm{Gal}$, is allowed to vary due to the discrepancy between $N_\mathrm{H}^\mathrm{Gal}=2.9\times10^{20}\mathrm{cm}^{-2}$ \citep[][]{2016HI4PI} and $N_\mathrm{H}^\mathrm{Gal}=3.6\times10^{20}\mathrm{cm}^{-2}$ \citep[NHtot tool,][]{2013Willingale}. The choice of the solar abundance calculated by \citet{2009Lodders} instead of \citet{2000Wilms} is to keep consistent with the subsequently used photoionization models in Sec.\ref{subsec:scan}, although it does not affect our conclusions (only a $\Delta\chi^2\sim10$ difference around O K-edge, $\sim0.53$\,keV, region).
Instead of using the high-density relativistic reflection model adopted in \citet{2018Mallick}, here we choose the warm Comptonization model plus a standard relativistic reflection component, of which the disk density is fixed at $\log(n_\mathrm{e}/\mathrm{cm}^{-3})=15$, in our analysis. It is because we find the fitting of the relativistic reflection model is much poorer ($\Delta\chi^2\sim670$) than the warm Comptonization scenario when we include the RGS data, probably due to a thick inner disk distorted by strong radiation pressure, breaking the thin-disk assumption of the reflection model. The seed photon of the warm Comptonization is fixed at a disk temperature of $10\,$eV , which is the value obtained by including the OM data (see Sec.\ref{subsec:scan}). 

\begin{figure}
	\includegraphics[width=0.5\textwidth, trim={50 50 30 20}]{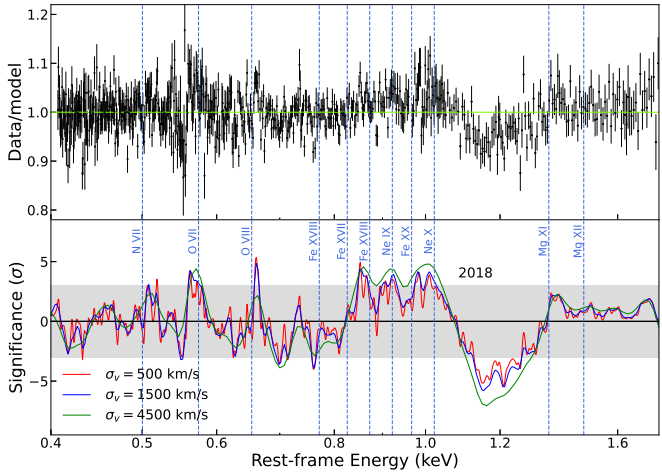}
    \caption{The data/model ratio ({\textit{upper}}) and single trial significance ({\textit{lower}}) obtained from the Gaussian line scan with different line widths (500, 1500, 4500 km/s) over the rest-frame stacked 2018 spectrum in RGS band. The vertical dashed lines represent the rest-frame positions of the known ion transitions as a reference. The grey region marks the significance of $3\sigma$.}
    \label{fig:ratio-gausian-2018}
\end{figure}

The fitted parameters of the stacked 2018 spectrum are listed in the third column of Tab.\ref{tab:fits}. The data/model ratio in the RGS band is shown in the upper panel of Fig.\ref{fig:ratio-gausian-2018}, featuring a broad absorption feature above 1\,keV. The results reveal a primary continuum with a slope of $\Gamma=2.26^{+0.01}_{-0.01}$ and a plasma temperature above $>196$\,keV, a warm Comptonization characterized by a temperature of $0.23^{+0.01}_{-0.01}$\,keV and a soft photon index $\Gamma_\mathrm{WC}=2.52^{+0.06}_{-0.06}$, and a relativistic reflection component with a reflection ratio of $f_\mathrm{Refl}=0.19^{+0.03}_{-0.02}$. The corresponding optical depth of the warm corona is $\tau^\mathrm{WC}_\mathrm{e}\sim30$ \citep{1985Zdziarski}. The spin of the black hole cannot be constrained and is thus fixed at $a_\star=0.998$. The inner radius of the disk is within $R_\mathrm{in}<23\,R_\mathrm{ISCO}$, where $R_\mathrm{ISCO}$ is the innermost stable circular orbit (ISCO). The inclination angle, ionization parameter, and iron abundance of the accretion disk are derived to be $i=34^{+1}_{-2}$ (deg), $\log(\xi/\mathrm{erg\,cm\,s}^{-1})=3.4^{+0.2}_{-0.1}$, and $A_\mathrm{Fe}=3.6^{+0.5}_{-0.6}$ (in units of solar abundance), respectively. The hot corona, if assumed in a lamppost geometry, is measured at the height of $h=47^{+26}_{-25}\,R_\mathrm{Horizon}$ above the accretion disk, where $R_\mathrm{Horizon}$ is the vertical event horizon of the Kerr black hole. The marginal differences between our results and \citet{2018Mallick} based on the archival 2013 observation ($i=46.4^{+1.9}_{-5.0}$ deg, $\log(\xi/\mathrm{erg\,cm\,s}^{-1})=2.96^{+0.04}_{-0.11}$, $A_\mathrm{Fe}=2.2^{+0.5}_{-0.6}$ in their fit) may come from the different explanations for the soft excess and the intrinsic variability of the source.
We apply the best-fit model to time-/flux-resolved spectra as well, with several invariable properties (i.e. $N_\mathrm{H}^\mathrm{Gal}$, $i$ and $A_\mathrm{Fe}$) on short-term timescales linked to those of the 2018 results. The results are listed in Tab.\ref{tab:fits}. There is no significant change in the broadband continuum during the 2018 observations, T1, T2, T3, within their uncertainties, confirming the prerequisite of the flux-resolved spectroscopy. The spectral slopes derived in flux-resolved spectra verify the softer-when-brighter behavior observed in Fig.\ref{fig:lc}.

\begin{table*}
\setlength{\tabcolsep}{3pt}
\renewcommand{\arraystretch}{1.2}
\centering
\caption{Best-fit parameters of the model {\tt tbabs*zashift*xabs\_xs*(nthComp+relxilllpCp+pion\_xs)} to the stacked 2018, time- (T) and flux-resolved (F) spectra. 
}
\begin{tabular}{lcccccccccc}
\hline
\hline
Description & Parameter & 2018 & T1 & T2 & T3 & T4 & F1& F2& F3\\
\hline
{\tt tbabs} &    $N^\mathrm{Gal}_\mathrm{H}$ ($10^{20}$ cm$^{-2}$)      & \multicolumn{7}{c}{$4.09^{+0.03}_{-0.05}$} \\
\hline
{\tt zashift} &    $z_\mathrm{Mrk 1044}$  & \multicolumn{7}{c}{$0.016^{\star}$} \\
\hline
{\tt nthComp}      & $\Gamma_\mathrm{WC}$  &  $2.52^{+0.06}_{-0.06}$    & $2.49^{+0.04}_{-0.05}$  & $2.51^{+0.05}_{-0.04}$ & $2.51^{+0.07}_{-0.06}$ & $2.42^{+0.13}_{-0.26}$ & $2.57^{+0.05}_{-0.05}$ & $2.50^{+0.05}_{-0.06}$ & $2.56^{+0.05}_{-0.02}$\\
                & $kT_\mathrm{e}$ (keV)  &  $0.23^{+0.01}_{-0.01}$ &   $0.23^{+0.01}_{-0.01}$  & $0.23^{+0.01}_{-0.01}$  & $0.22^{+0.01}_{-0.01}$ & $0.20^{+0.02}_{-0.02}$ & $0.24^{+0.01}_{-0.01}$ & $0.22^{+0.01}_{-0.01}$ & $0.25^{+0.01}_{-0.01}$\\
                &  $N_\mathrm{WC}$ ($10^{-3}$)     &     $6.0^{+0.4}_{-0.5}$      & $7.5^{+0.7}_{-0.6}$   & $5.5^{+0.3}_{-0.4}$  & $5.3^{+0.4}_{-0.5}$ & $3.7^{+0.4}_{-0.5}$ &  $4.6^{+0.4}_{-0.2}$ & $5.9^{+0.3}_{-0.5}$ & $9.3^{+0.5}_{-0.6}$\\
\hline
{\tt relxilllpCp}   & $h$ ($R_\mathrm{Horizon}$)  & $47^{+26}_{-25}$&   $>46$ & $>40$                        &$33^{+21}_{-28}$ & $-17^{+18}_{-12}$ & $>37$ & $26^{+28}_{-20}$  & $>24$ \\
                    & $a_\star$ ($cJ/GM^2$) & \multicolumn{7}{c}{$0.998^{\star}$}\\
                    & $i$ (deg) & \multicolumn{7}{c}{$34^{+1}_{-2}$} \\	
                    & $R_\mathrm{in}$ (${R_\mathrm{ISCO}}$) & $<23$  & $<82$ & $<97$ &$<30$ & $15^{+11}_{-11}$ & $<49$ & $<29$  & $<42$ \\	
                    & $\Gamma$   &  $2.26^{+0.01}_{-0.01}$ &  $2.29^{+0.01}_{-0.03}$  & $2.23^{+0.01}_{-0.02}$ & $2.23^{+0.02}_{-0.02}$ & $2.22^{+0.03}_{-0.03}$ & $2.18^{+0.01}_{-0.02}$ & $2.27^{+0.02}_{-0.02}$  & $2.31^{+0.02}_{-0.02}$ \\
                    & $\log(\xi/\mathrm{erg\,cm\,s^{-1})}$ & $3.4^{+0.2}_{-0.1}$  & $3.6^{+0.2}_{-0.3}$ & $3.4^{+0.1}_{-0.2}$ &$3.3^{+0.1}_{-0.1}$ &$3.2^{+0.2}_{-0.2}$ & $3.3^{+0.1}_{-0.1}$ & $3.3^{+0.2}_{-0.1}$  & $2.1^{+0.4}_{-0.2}$ \\
                    &  $A_{\mathrm{Fe}}$ & \multicolumn{7}{c}{$3.6^{+0.5}_{-0.6}$} \\
                    & $kT_\mathrm{e}$ (keV) & $>196$  & $>51$ & $>30$ &$>23$ &$>18$ &$>38$ &$>25$  & $>21$ \\
                    & $f_\mathrm{Refl}$ &  $0.19^{+0.03}_{-0.02}$  &   $0.23^{+0.05}_{-0.05}$  & $0.20^{+0.04}_{-0.05}$ &$0.29^{+0.06}_{-0.06}$ & $0.35^{+0.14}_{-0.09}$ & $0.25^{+0.03}_{-0.05}$ & $0.28^{+0.05}_{-0.05}$  & $0.32^{+0.10}_{-0.09}$ \\
                    &  $N_\mathrm{refl}$ ($10^{-5}$)     & $9.4^{+0.8}_{-0.2}$  & $9.9^{+1.0}_{-1.1}$  & $7.8^{+0.9}_{-0.3}$ & $8.8^{+14.0}_{-0.8}$ &$8^{+11}_{-1}$ & $6.9^{+5.0}_{-0.3}$ & $10^{+8}_{-9}$  & $13^{+17}_{-1}$ \\
\hline
 broadband    &  $\chi^2$/d.o.f. & 1319/733  &   987/731 &  922/731& 956/731 &750/730 & 950/734 & 933/734 &939/733   \\
\hline
{\tt xabs\_xs}     & $N_\mathrm{H}$ ($10^{21}$\,cm$^{-2}$)   &  $2.3^{+0.5}_{-0.4}$   & $2.2^{+3.3}_{-0.4}$ & $2.0^{+1.2}_{-0.4}$ &$1.9^{+1.1}_{-0.2}$ & $0.04^{+0.02}_{-0.02}$ &$1.8^{+1.0}_{-0.3}$ & $2.1^{+1.7}_{-0.6}$  & $5.4^{+4.0}_{-3.2}$ \\
        &    $\log(\xi/\mathrm{erg\,cm\,s^{-1})}$   &  $3.72^{+0.08}_{-0.10}$ & $3.73^{+0.23}_{-0.09}$ & $3.74^{+0.18}_{-0.14}$ &$3.74^{+0.30}_{-0.09}$ & $2.01^{+0.31}_{-0.25}$ & $3.74^{+0.17}_{-0.11}$ &$3.75^{+0.12}_{-0.20}$  & $4.0^{+0.1}_{-0.2}$ \\
        &    $\sigma_\mathrm{v}$ (km/s) &  $11800^{+4600}_{-4000}$  & $8600^{+5400}_{-4400}$ & $9500^{+5900}_{-6800}$ & $8500^{+4400}_{-3900}$ & $1000^{+1000}_{-800}$  & $9000^\star$ &$9000^\star$  & $9000^\star$ \\        
        &    $z_\mathrm{LOS}$  &  $-0.153^{+0.008}_{-0.016}$  & $-0.181^{+0.010}_{-0.007}$ & $-0.145^{+0.015}_{-0.021}$ & $-0.143^{+0.010}_{-0.012}$ & $-0.082^{+0.002}_{-0.002}$ & $-0.146^{+0.013}_{-0.018}$ &$-0.179^{+0.010}_{-0.018}$  & $-0.188^{+0.013}_{-0.007}$ \\
\hline
  broadband+abs      &  $\chi^2$/d.o.f. & 1183/729 &  925/727 &  884/727                           & 928/727& 724/726 & 915/731 &887/731 &895/730 \\
\hline
{\tt pion\_xs}     & $N_\mathrm{H}$ ($10^{20}$\,cm$^{-2}$)   &  $2.1^{+0.6}_{-0.5}$   & $1.2^{+1.3}_{-0.7}$ & $8^{+7}_{-6}$ &$2.4^{+0.4}_{-1.4}$ & $2.7^{+0.7}_{-1.1}$ &$0.3^{+0.1}_{-0.1}$ & $2.6^{+0.9}_{-1.8}$  & $2.4^{+0.7}_{-1.4}$ \\
        &    $\log(\xi/\mathrm{erg\,cm\,s^{-1})}$   &  $2.5^{+0.1}_{-0.1}$ & $2.4^{+0.3}_{-0.3}$ & $3.0^{+0.1}_{-0.4}$ &$2.3^{+0.1}_{-0.2}$ & $2.6^{+0.1}_{-0.2}$ & $1.6^{+0.6}_{-0.2}$ &$2.6^{+0.1}_{-0.3}$  & $2.4^{+0.2}_{-0.1}$ \\
        &    $\sigma_\mathrm{v}$ (km/s) &  $1500^{+800}_{-700}$  & $1500^{\star}$ & $1000^{+800}_{-750}$ & $2200^{+3400}_{-1300}$ & $1500^{\star}$ & $800^{+750}_{-500}$ &$<3300$  & $1300^{+1200}_{-800}$ \\
        &    $z_\mathrm{LOS}$ ($10^{-2}$) &  $-1.1^{+0.2}_{-0.1}$  & $-1.5^{+0.7}_{-0.5}$ & $-1.2^{+0.2}_{-0.2}$ & $-1.1^{+0.3}_{-0.4}$ & $-1.1^{+0.3}_{-0.4}$ & $-0.9^{+0.1}_{-0.2}$ &$-1.3^{+0.4}_{-0.5}$  & $-1.5^{+0.3}_{-0.3}$ \\
\hline
  broadband+abs+em      &  $\chi^2$/d.o.f. & 1135/725 &      913/724 &  853/723  & 886/723& 698/723 & 877/728 &867/728 &864/727 \\
\hline
Flux (0.4-10\,keV) &$F$ ($10^{-11}$\,erg/cm$^2$/s) & $6.34^{+0.03}_{-0.04}$ & $7.38^{+0.07}_{-0.06}$ & $5.60^{+0.09}_{-0.08}$& $5.90^{+0.12}_{-0.07}$ & $4.56^{+0.06}_{-0.07}$ & $4.92^{+0.09}_{-0.08}$& $6.45^{+0.10}_{-0.10}$ & $8.59^{+0.09}_{-0.11}$\\
\hline
\end{tabular}
\label{tab:fits}
\begin{flushleft}
{$^{\star}$ The parameter is fixed.}
\end{flushleft}
\vspace{-4mm}
\end{table*}

\subsection{Gaussian Line Scan}\label{subsec:gaussianscan}
To better visualize and identify the atomic features upon the continuum, we launch a blind Gaussian line scan over the spectra. We fit an additional Gaussian line with a logarithmic grid of energy steps over $0.4\mbox{--}10$\,keV band upon the continuum model and record the $\Delta\chi^2$ improvements. The energy centroid and the line width are fixed at each step, while the normalization is free. We adopt three line widths $\sigma_v$ of 500, 1500, 4500 km/s and the corresponding numbers of energy steps are 2000, 700, 300, respectively, in order to match the RGS spectral resolution power ($R_\mathrm{RGS}\sim 150\mbox{--}800$). 

The scan results provide a rough estimate of the single trial detection significance of each Gaussian line, in the form of the square root of $\Delta\chi^2$ times the sign of the normalization \citep{1979Cash}. The scan results over the 2018 spectrum in the RGS band are shown in the bottom panel of Fig.\ref{fig:ratio-gausian-2018}. The rest-frame energies of the known strong ionic transition lines in the soft X-ray band are marked by the vertical blue \textit{dashed} lines. We identify the O \textsc{vii} and O \textsc{viii} emission lines close to their rest-frame positions as well as several emission features in the Ne \textsc{ix/x} and Fe \textsc{xvii-xx} region. No absorption features are found at/close to their rest frames. The strongest absorption feature is located around 1.2\,keV with a broad line width, likely from blueshifted Fe and Ne ionic absorption lines.
\begin{figure}
	\includegraphics[width=0.5\textwidth, trim={50 30 30 20}]{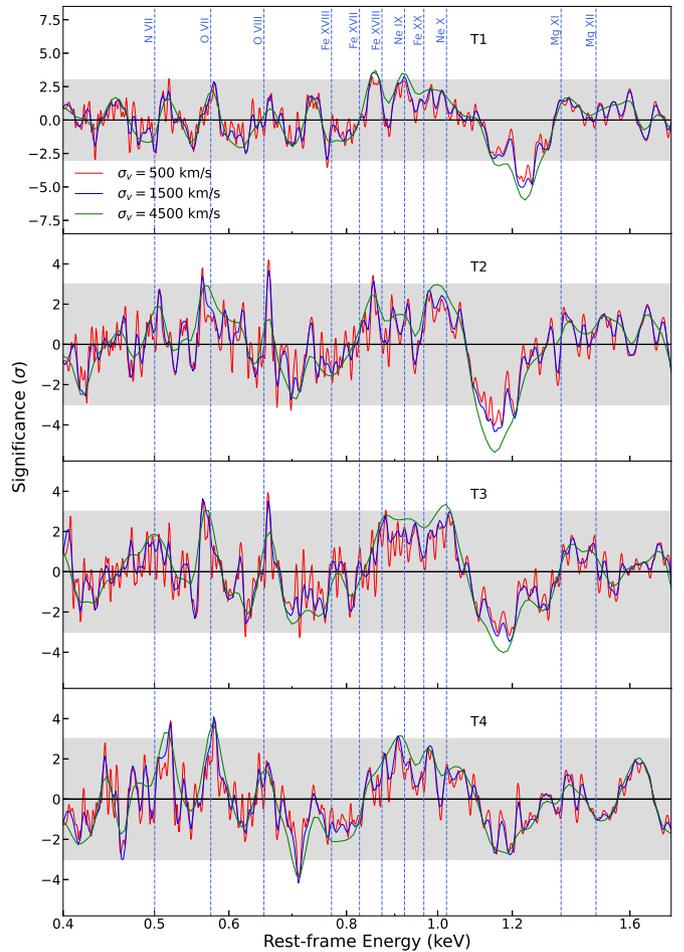}
    \caption{Similar to the bottom panel of Fig.\ref{fig:ratio-gausian-2018} but the scan is performed on the time-resolved spectra.}
    \label{fig:gaussian_T}
\end{figure}
\begin{figure}
	\includegraphics[width=0.5\textwidth, trim={50 30 30 20}]{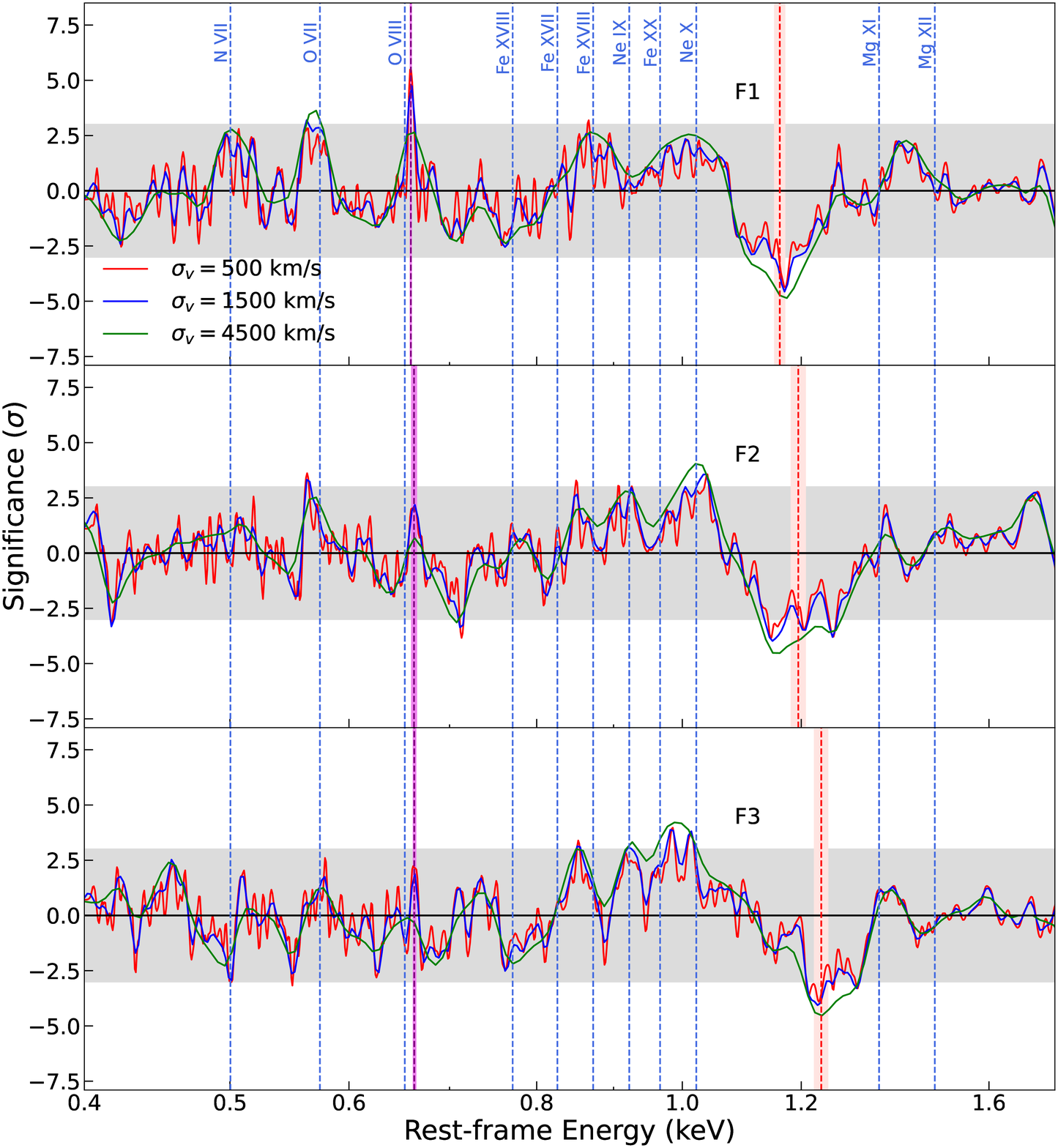}
    \caption{Similar to the bottom panel of Fig.\ref{fig:ratio-gausian-2018} but the scan is performed on the flux-resolved spectra. The vertical dashed red/purple lines and the red/purple shadowed areas indicate the position of the centroid of the absorption/emission feature and the corresponding uncertainty. (see Tab.\ref{tab:gaussian}).}
    \label{fig:gaussian_F}
\end{figure}

The same approach is then applied to the time-/flux-resolved spectra with the same primary settings, of which results are shown in Fig.\ref{fig:gaussian_T} and Fig.\ref{fig:gaussian_F}, separately. There are no significant differences among the results of time-resolved spectra, except for the T4 spectrum. Due to its low flux, the T4 result has a weaker detection significance of lines and the strongest absorption feature becomes around 0.7\,keV, suggesting a different ionization state of the absorber in T4 observation. In the other three spectra, the emission (O \textsc{vii}, O \textsc{viii} and $0.8\mbox{--}1\,$keV emission) and absorption ($\sim1.2$\,keV line) features observed in the 2018 spectrum are all obvious. The absorption feature in T1 spectrum, the brightest observation, seems to be more blueshifted than others. In addition, the line width of the 1.2\,keV trough is not as broad as that of the 2018 spectrum, due to the stack effect of the variable feature. Comparing the flux-resolved results, we notice a decreasing significance of the O \textsc{vii} and O \textsc{viii} lines and an increasing blueshift of the absorption feature with the source luminosity, implying the existence of possible wind-luminosity relations. We thus fit the absorption feature with a \texttt{Gaussian} model and the parameters are listed in Tab.\ref{tab:gaussian}. The energy centroid of the absorption feature increases with the flux and is highlighted in \textit{red} in Fig.\ref{fig:gaussian_F}. The best-fit parameters for the O \textsc{viii} emission line are also listed in Tab.\ref{tab:gaussian} and depicted in \textit{purple} in Fig.\ref{fig:gaussian_F}, indicating a slightly increasing blueshift.

\begin{table}
\centering
\caption{Best-fit parameters of an additional \texttt{Gaussian} model over the continuum model for the absorption feature around 1.2 keV and the O \textsc{viii} emission line. 
}
\begin{tabular}{lcccccc}
\hline
\hline
Parameter & F1 & F2  & F3  \\
\hline
$\mathrm{E}_\mathrm{rest}^\mathrm{abs}$ (keV) & $1.160^{+0.015}_{-0.016}$ & $1.193^{+0.024}_{-0.021}$ & $1.239^{+0.022}_{-0.024}$ \\
$\mathrm{EW}_\mathrm{abs}$ (eV) &  $50^{+19}_{-13}$ & $74^{+24}_{-19}$ & $73^{+30}_{-22}$\\
$\Delta\chi^2_\mathrm{abs}$ & 40 & 45 & 39 \\
\hline
$\mathrm{E}_\mathrm{rest}^\mathrm{em}$ (keV) & $0.660^{+0.001}_{-0.001}$ & $0.663^{+0.004}_{-0.004}$ & $0.663^{+0.002}_{-0.001}$ \\
$\mathrm{EW}_\mathrm{em}$ (eV) &  $1.4^{+1.0}_{-0.7}$ & $<6.4$ & $2.2^{+1.5}_{-1.3}$\\
$\Delta\chi^2_\mathrm{em}$ & 27 & 7 & 10 \\
\hline
\end{tabular}
\label{tab:gaussian}
\end{table}

\subsection{Search for outflows}\label{subsec:scan}
To study the emission/absorption lines discovered in Sec.\ref{subsec:gaussianscan}, we employ the physical photoionization model, \texttt{pion}, in the SPEX package \citep{1996Kaastra}. This code self-consistently calculates the photoionization equilibrium and synthetic spectra of the gas irradiated by a given radiation field.

\begin{figure}
	\includegraphics[width=0.45\textwidth, trim={20 10 20 10}]{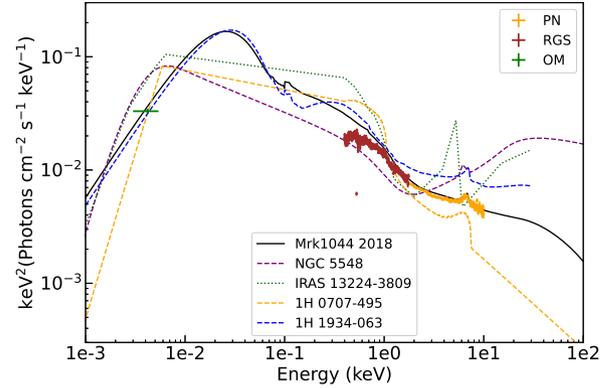}
    \caption{The averaged spectral energy distribution of Mrk 1044 in 2018 compared with other Seyfert galaxies (NGC 5548, \citealt{2015Mehdipour}; IRAS 13224-3809, \citealt{2018Jiang}; 1H 0707-495, \citealt{2021Xu}; 1H 1934-063, \citealt{2022Xu}). The EPIC-pn, RGS, and OM data are shown as well, where the deviations from the best-fit SED come from the removal of the Galactic absorption, redshift, and dust-reddening components.} 
    \label{fig:SED}
\end{figure}

The intrinsic spectral energy distribution (SED) of Mrk 1044 inputted into \texttt{pion} is derived from the UV to hard X-ray energies. Due to the stability of the OM flux, we stack the OM spectra and model it with an additional \texttt{diskbb} component, characterized by a temperature of $10^{+21}_{-6}$\,eV. Such a temperature is relatively low for the accretion disk around an SMBH with a mass of $\sim3\times10^6\,M_\odot$ \citep{1973Shakura}, might be explained by the truncated disk (suggested by the inner radii $R_\mathrm{in}$ in \texttt{relxilllpCp}, see Tab.\ref{tab:fits}). The interstellar extinction \citep[$E_\mathrm{B-V}=0.031$,][]{2016Marinello} is also considered. The SED of Mrk 1044 in 2018 is shown in Fig.\ref{fig:SED} compared with other Seyfert galaxies, where it shares a similar soft SED with 1H 1934-063. The observed data are shown on top of the SED, where the deviations from the best-fit SED come from the removal of the Galactic absorption, redshift, and dust-reddening components. By measuring the bolometric luminosity ($10^{-3}\mbox{--}10^{3}$\,keV) predicted by the model, $L_\mathrm{Bol}\sim1.4\times10^{44}\,$erg/s, we thus estimate the Eddington ratio of Mrk 1044 at $\lambda_\mathrm{Edd}=L_\mathrm{Bol}/L_\mathrm{Edd}\sim0.4$, adopting a SMBH mass of $2.8\times10^{6}\,M_\odot$ \citep{2015Du}, where  $L_\mathrm{Edd}=4\pi GM_\mathrm{BH}m_\mathrm{p}c/\sigma_\mathrm{T}$ is the Eddington luminosity. Although our estimated Eddington ratio is slightly different from the literature \citep[$\lambda_\mathrm{Edd}=0.59$,][]{2010Grupe}, due to the different masses adopted \citep[$M_\mathrm{BH}=2.1\times10^{6}\,M_\odot$,][]{2010Grupe}, it still implies a high-accretion system and the value is comparable to that of 1H 1934-063 \citep[$\lambda_\mathrm{Edd}=0.40^{+0.91}_{-0.27}$,][]{2022Xu} calculated from the same approach.

To take advantage of both the advanced reflection model (RELXILL), implemented in XSPEC, and the \texttt{pion} model in SPEX, we adopt the code used in \citet{2019Parker} to construct the tabulated model, which is an XSPEC version of \texttt{pion}, called \texttt{pion\_xs}. In this paper, we only make use of the emission component of \texttt{pion} (i.e. solid angle $\Omega=1$, and covering fraction $C_\mathrm{F}=0$), while the absorption component is explained by \texttt{xabs\_xs}, transferred from \texttt{xabs} in SPEX ($C_\mathrm{F}=1$). The \texttt{pion} and \texttt{xabs} models are characterized by four main parameters, including the column density $N_\mathrm{H}$, the ionization parameter $\log\xi$, the turbulence velocity $\sigma_v$ and the line-of-sight (LOS) redshift of gas $z_\mathrm{LOS}$. 

\begin{figure*}
	\includegraphics[width=0.45\textwidth, trim={20 10 20 10}]{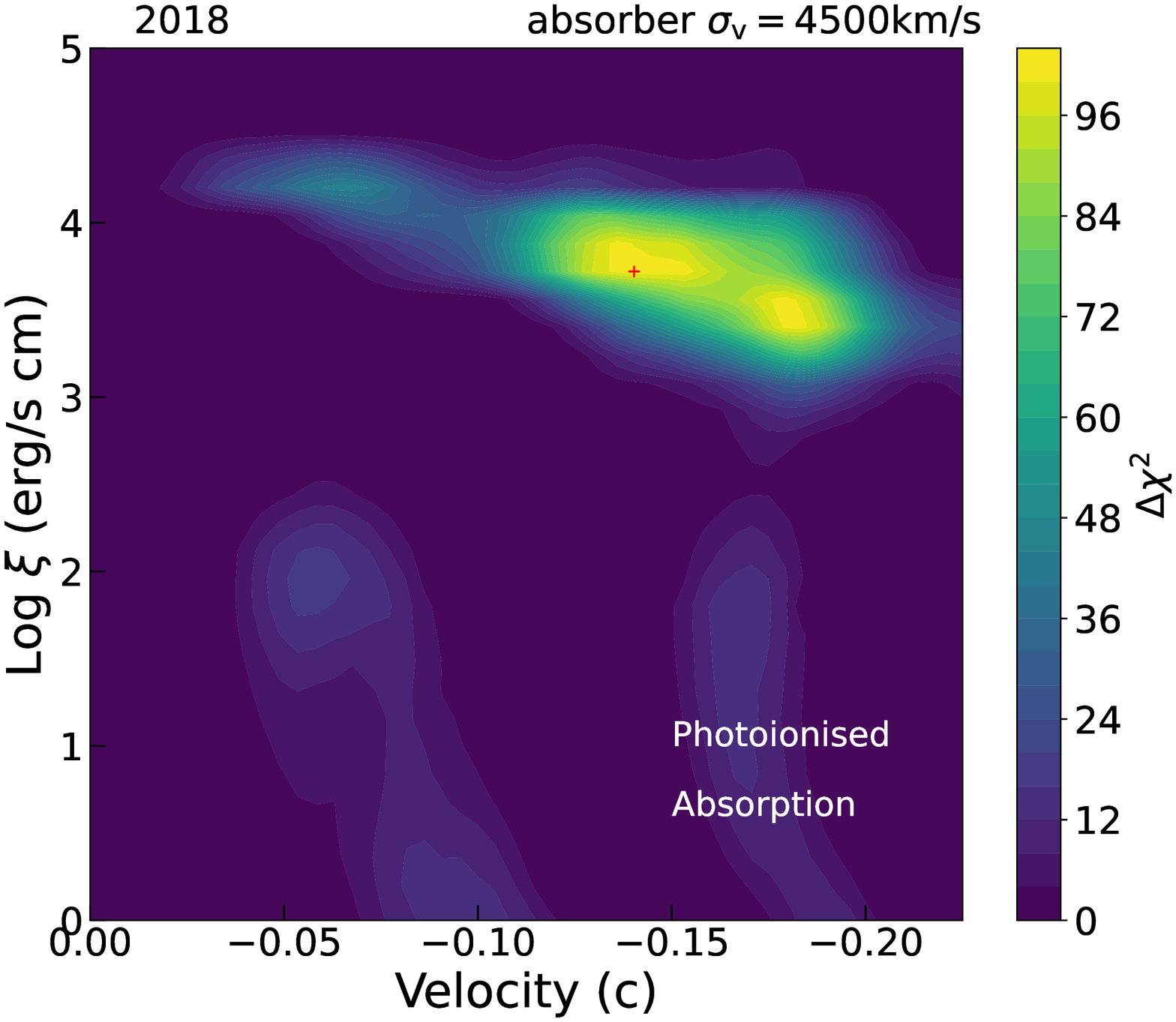}
 	\includegraphics[width=0.45\textwidth, trim={20 10 20 10}]{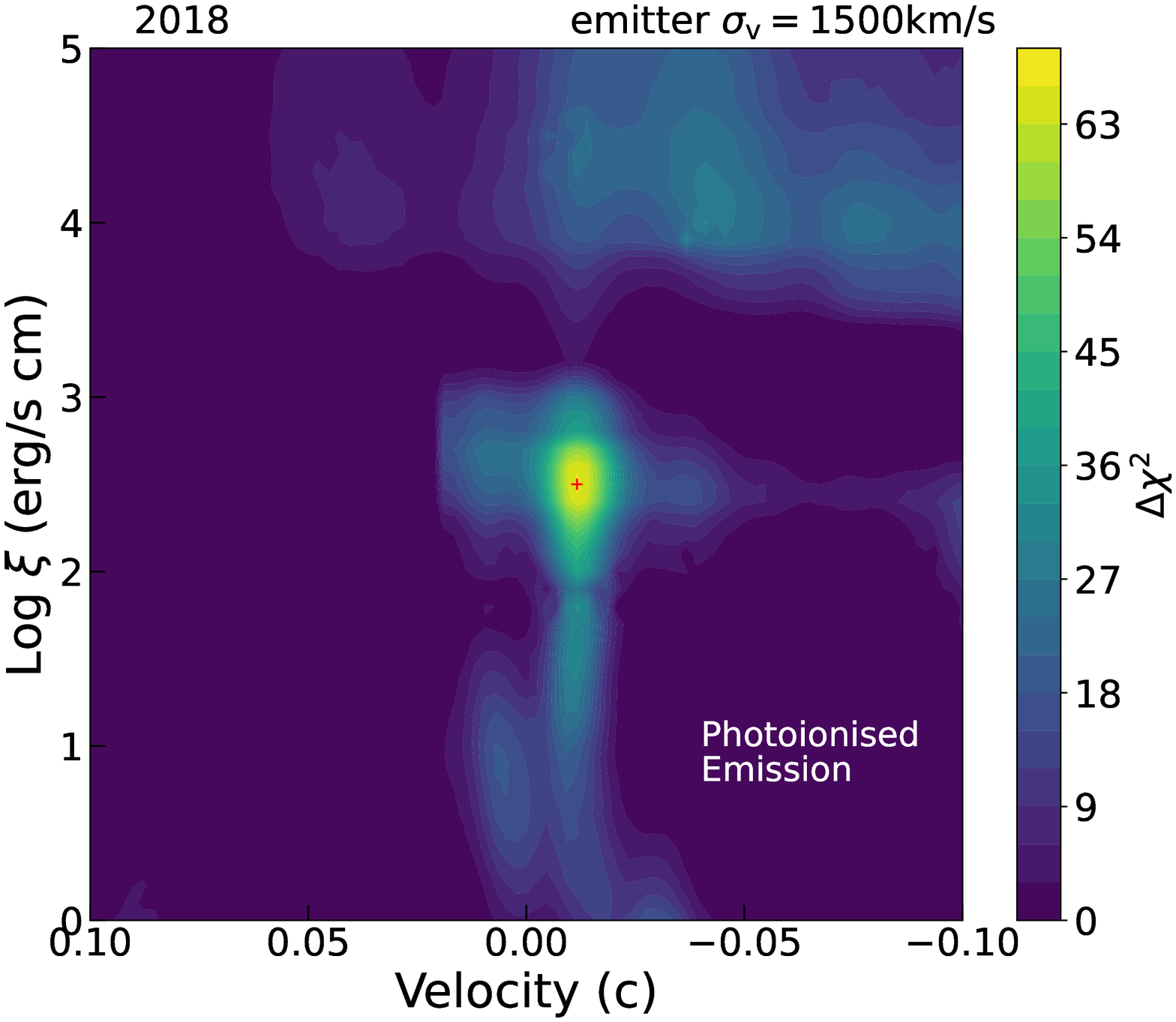}
    \caption{Photoionization absorption (\textit{left}) /emission (\textit{right}) model search for the stacked 2018 spectrum of Mrk 1044 over the broadband model. The color illustrates the statistical improvement after adding an absorption/emission component. The best-fit solution is marked by a red cross.}
    \label{fig:scan}
\end{figure*}
\begin{figure}
	\includegraphics[width=0.49\textwidth, trim={0 80 0 0}]{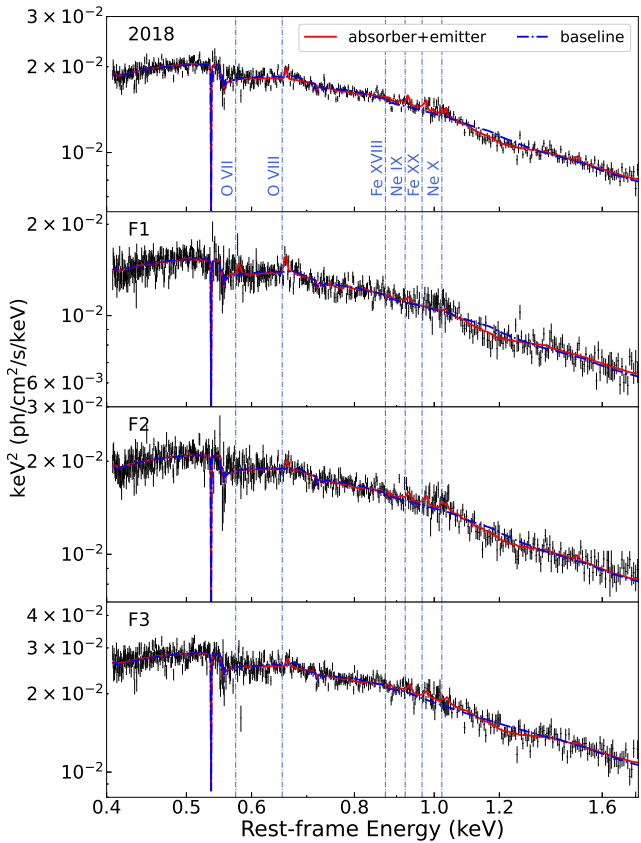}
    \caption{The stacked 2018 (\textit{first}) and flux-resolved (from \textit{second} to \textit{fourth}) RGS spectra (\textit{black} dots with errors) of Mrk 1044. Each panel contains the fits with the baseline continuum model (\textit{blue}) and the continuum plus a \texttt{pion\_xs} and \texttt{xabs\_xs} model (\textit{red}). The rest-frame energies of the relevant ion transitions are marked by the vertical dashed lines.}
    \label{fig:spectrum}
\end{figure}

\begin{figure*}
    \begin{tabular}{ll}
	\includegraphics[width=0.4\textwidth, trim={0 0 0 0},]{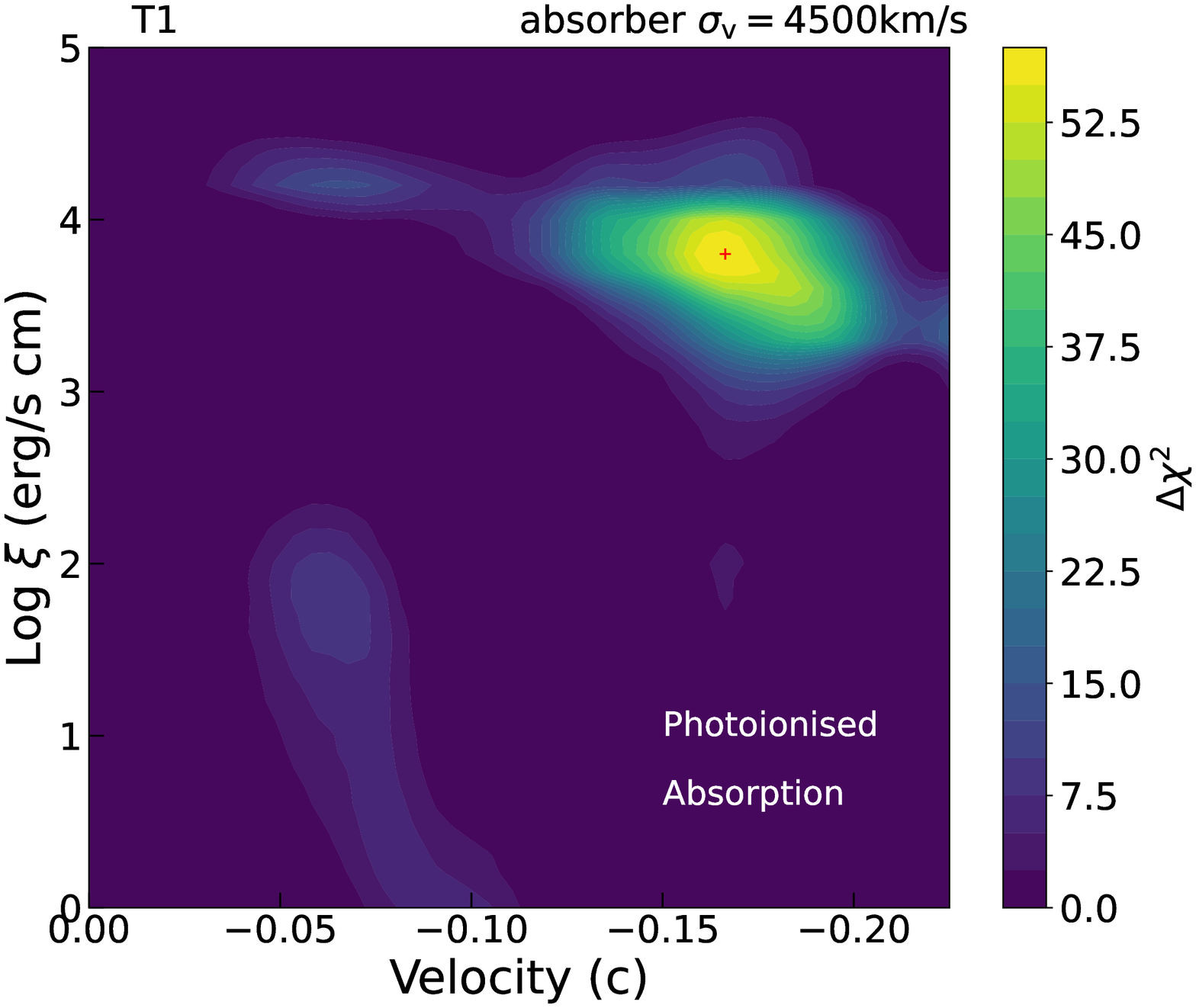}&
        \includegraphics[width=0.4\textwidth, trim={0 0 0 0},]{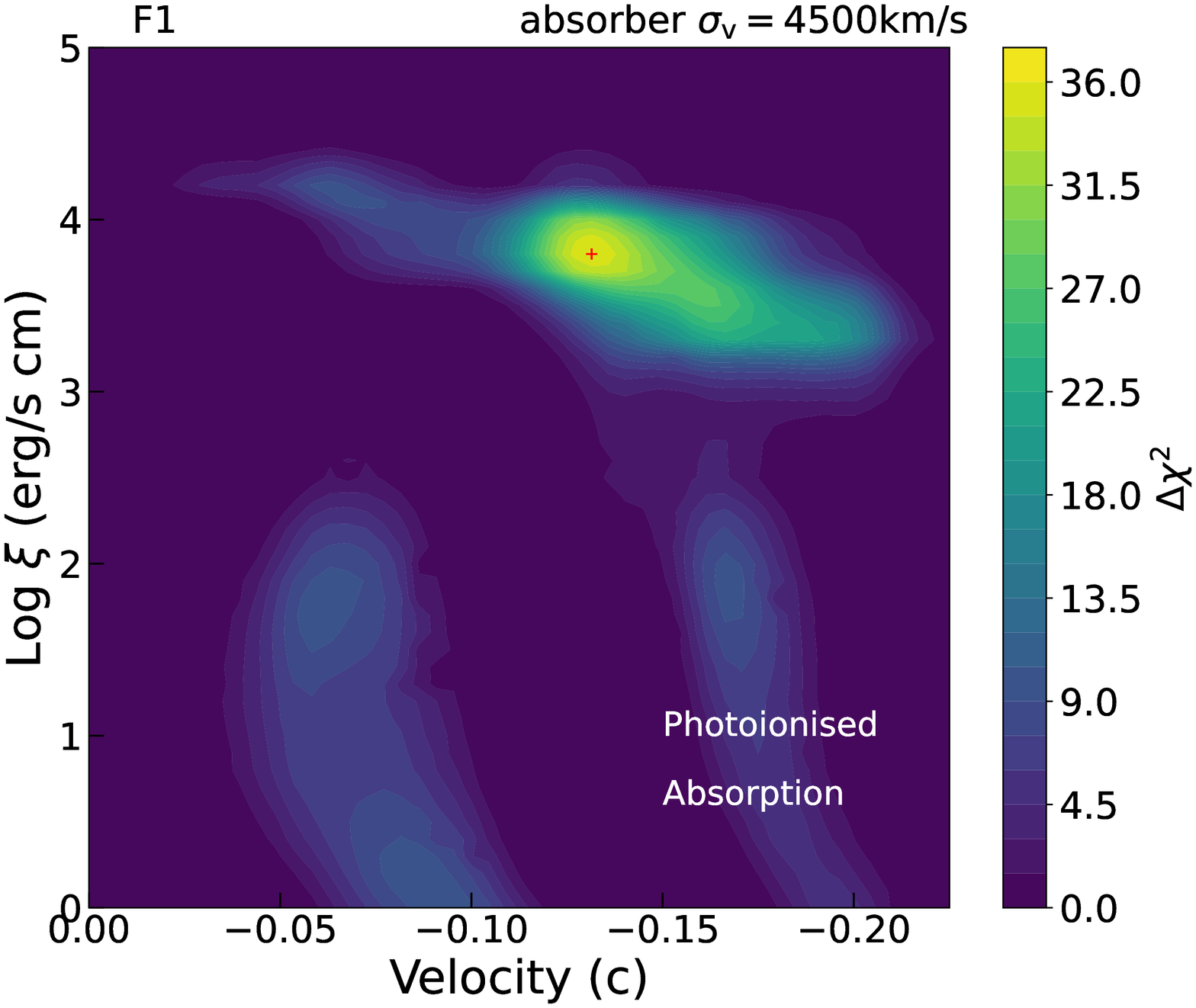}\\
	\includegraphics[width=0.4\textwidth, trim={0 0 0 0},]{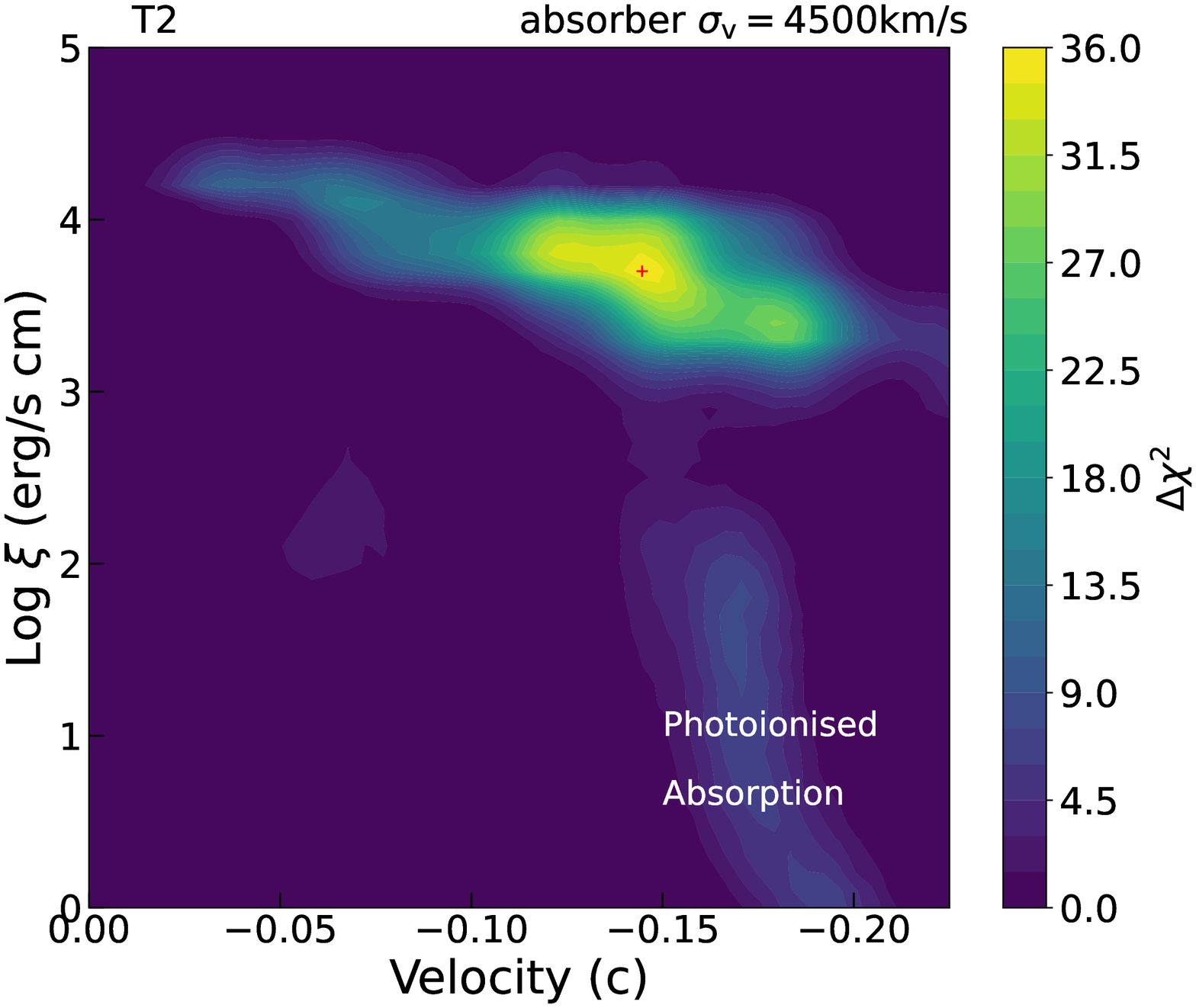}&
        \includegraphics[width=0.4\textwidth, trim={0 0 0 0},]{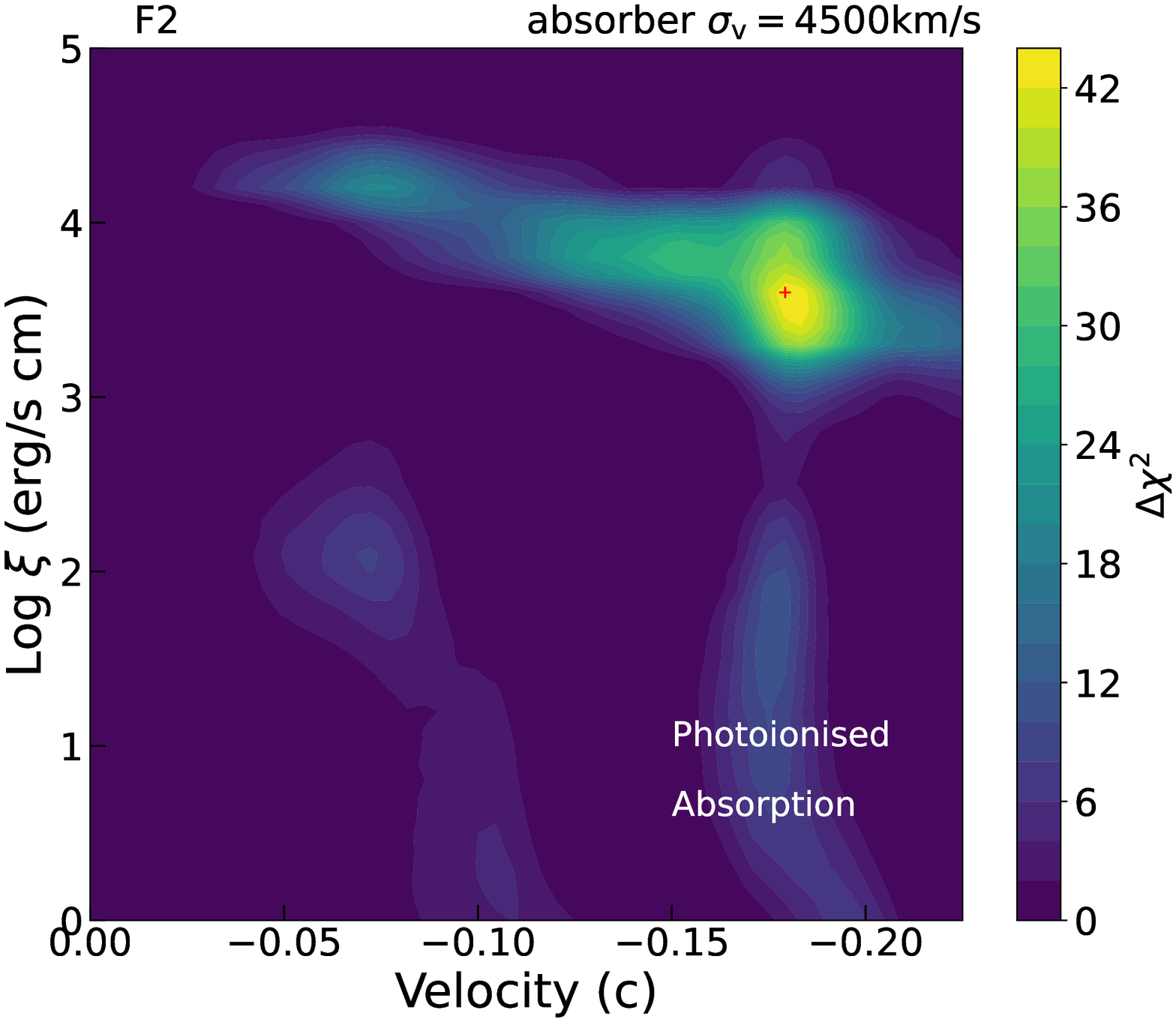}\\
	\includegraphics[width=0.4\textwidth, trim={0 0 0 0},]{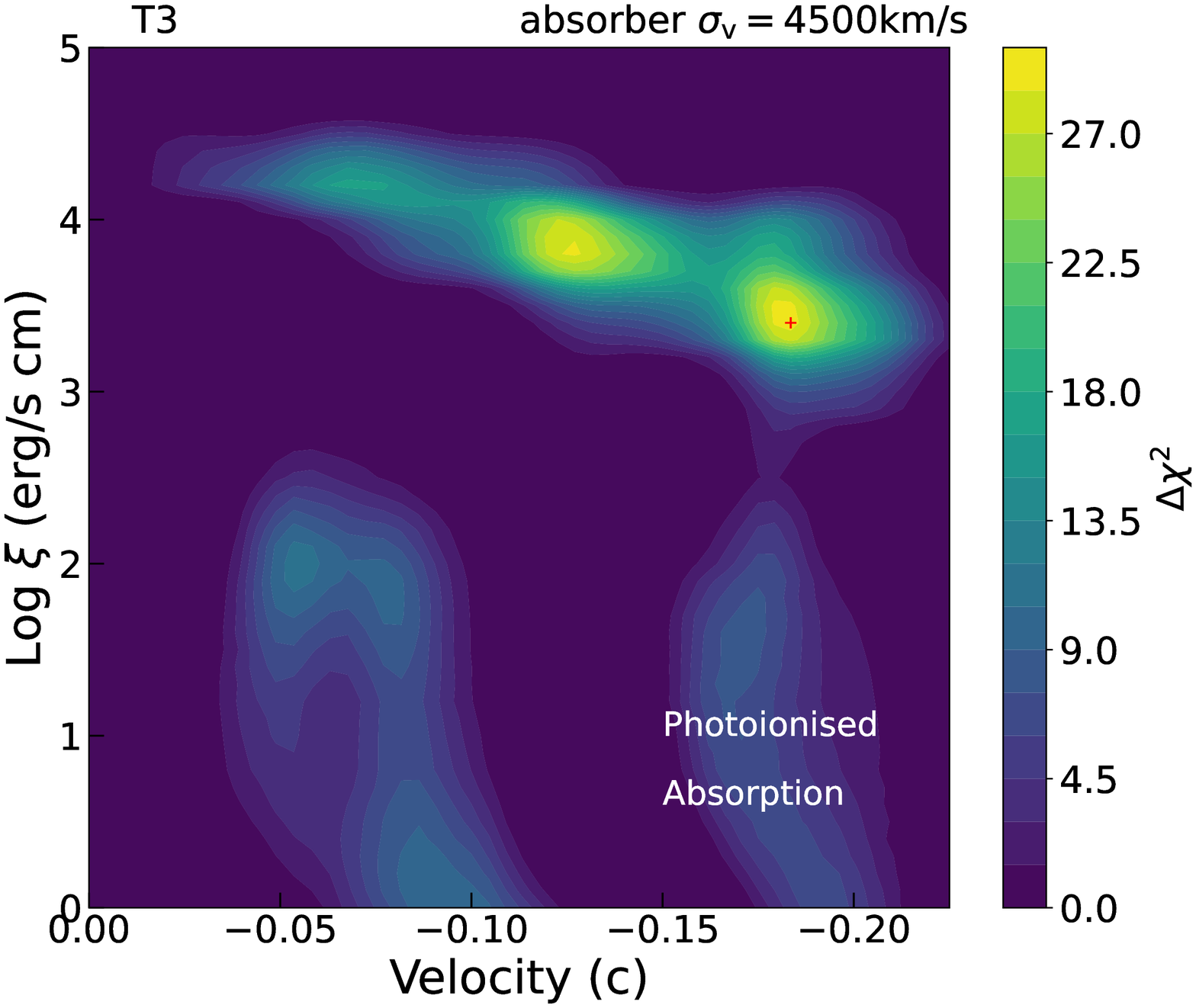}&
        \includegraphics[width=0.4\textwidth, trim={0 0 0 0},]{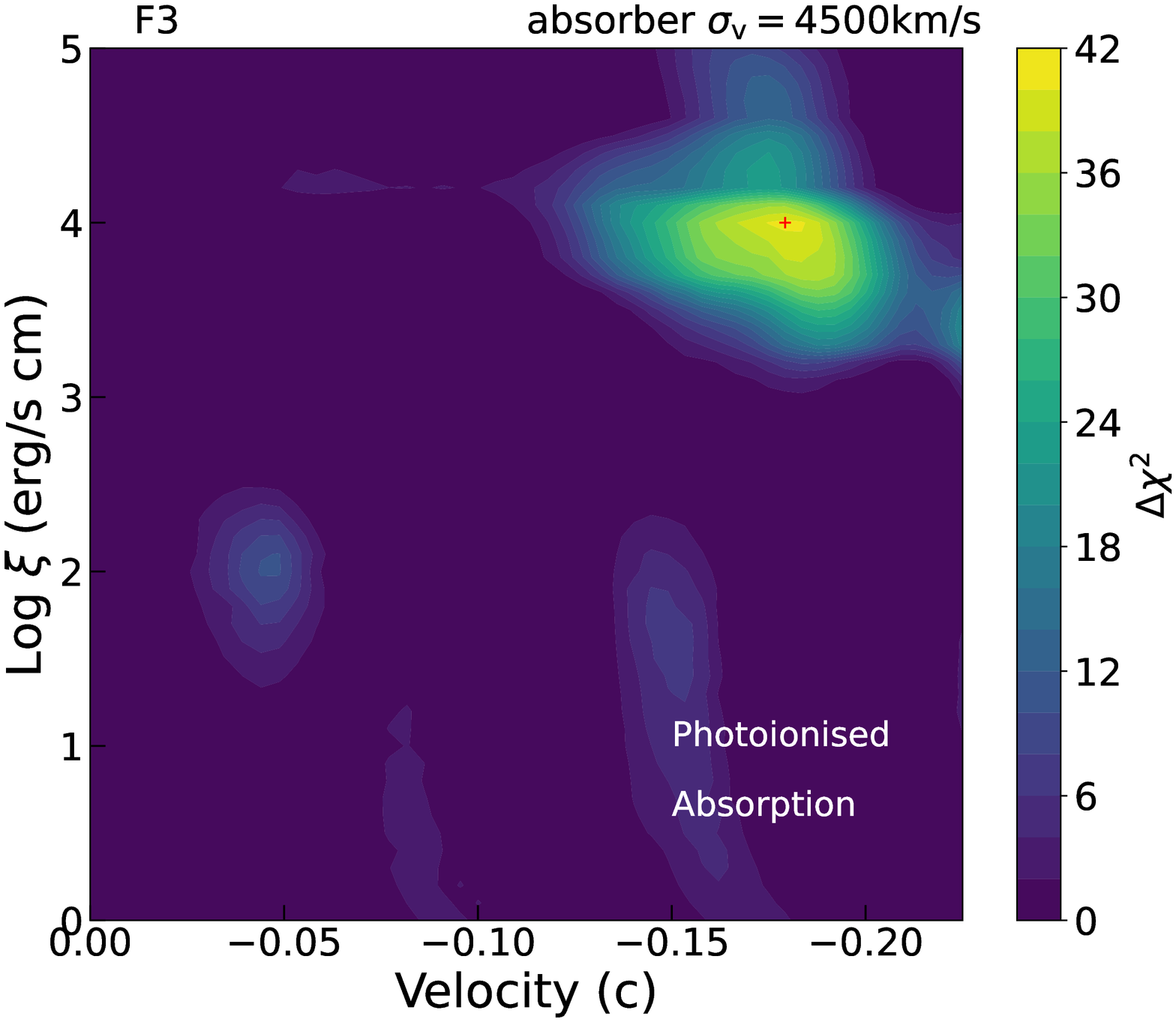}\\
	\includegraphics[width=0.4\textwidth, trim={0 0 0 0}]{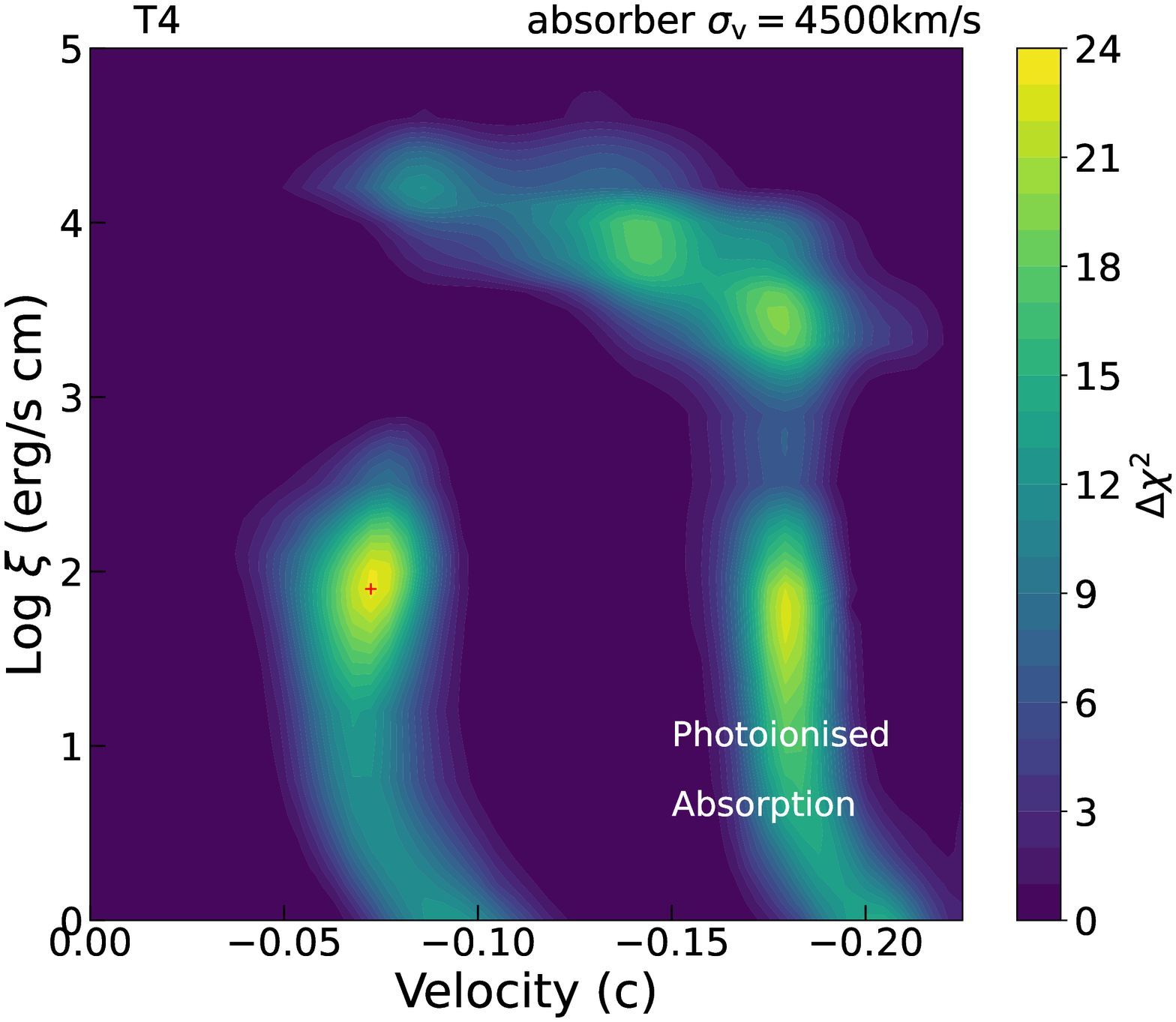}\tabularnewline
      \end{tabular} \tabularnewline

    \caption{Similar to the left panel of Fig.\ref{fig:scan} but the scan is performed on the time- (T, \textit{left}) and flux- (F, \textit{right}) resolved spectra.}
    \label{fig:scan_all}
\end{figure*}

\subsubsection{Absorption}\label{subsubsec:scanabsorption}
To locate the globally best-fit solution of the absorbing gas, we launch a systematic scan over a multi-dimension grid of the parameters ($\log\xi, z_\mathrm{LOS}, \sigma_v$) of \texttt{xabs\_xs}, following \citet{2021Xu,2022Xu}. The range of $\log\xi$ is $0\mbox{--}5$ with a step of $\Delta\log\xi=0.1$. The grid of the turbulent velocity $\sigma_v$ is the same as that of the Gaussian line scan ($\sigma_v=500, 1500, 4500$\,km/s). The LOS velocity, $z_\mathrm{LOS}$, ranges from -0.35 to 0, with an increment depending on the choice of $\sigma_v$ (c$\Delta z_\mathrm{LOS}=500, 700, 1500$\,km/s for $\sigma_v=500, 1500, 4500$\,km/s respectively). The scan is performed upon the best-fit model obtained in Sec.\ref{subsec:continuum}. The column density, $N_\mathrm{H}$, and continuum parameters are left free. The $\Delta\chi^2$-statistics improvement is recorded at each grid to reveal the detection significance of the absorbing gas. One advantage of the scan is to show the location of all possible solutions in the parameter space, probably revealing multiphase outflows.


The scan result of the 2018 spectrum is shown in the left panel of Fig.\ref{fig:scan}, where the best solution is marked with a red cross. Because of the consistent solutions with different turbulent velocities, we only present the result with $\sigma_v=4500$\,km/s in this paper, which has the largest detection significance. The velocity on the X-axis is the relativistically corrected velocity according to the equation: $v/c=\sqrt{(1+z_\mathrm{LOS})/(1-z_\mathrm{LOS})}-1$. It reveals a strong detection ($\Delta\chi^2=103$) of a highly ionized ($\log\xi=3.72$) and ultra-fast ($z_\mathrm{LOS}=-0.15$) absorber. If we allow the line width to vary, the solution of the direct fit ($\sigma_v\sim12000$\,km/s, $N_\mathrm{H}=2.3\times10^{21}\,\mathrm{cm}^{-2}$) is listed in Tab.\ref{tab:fits}, consistent with our scan result. The contribution of this absorber to the modeling is visible in the top panel of Fig.\ref{fig:spectrum}, mainly around 1.2\,keV from blueshifted Fe \textsc{xxii-xxiv} and Ne \textsc{x}, without any absorption features detected in the EPIC band probably due to its relatively low column density and the soft SED.

The same scan is also performed for the time-/flux-resolved spectra, shown in Fig.\ref{fig:scan_all}, and their best-fit solutions are summarized in Tab.\ref{tab:fits}. The UFO detection significance in each spectrum is at least $4\sigma$, i.e. $\Delta\chi^2=24.5$ for 4 degrees of freedom (d.o.f.). We also calculate the X-ray flux between 0.4 and 10 keV with the \texttt{cflux} model, presented in Tab.\ref{tab:fits}. The best-fit velocity of T3 is around $-0.2c$ with a narrow line width of $<178$\,km/s, which is quite different from the absorber in the other two consecutive observations. The scan plot reveals another degenerate region below $-0.15c$ with a broad line width of $\sim8500$\,km/s and comparable statistics ($\Delta\chi^2\sim3$). Therefore, to ensure we are tracing the same absorber, we adopt this slow and broad solution in our analysis. 

Among the time-resolved spectra, apart from T4, the ionization state and column density of the UFO are consistent within their uncertainties. The velocity of the UFO indicatively has an increasing trend with the source flux. As for T4, instead of the 1.2\,keV feature, the best-fit solution of T4 explains the blueshifted O \textsc{viii} line around 0.7\,keV (see the bottom panel of Fig.\ref{fig:gaussian_T}). It means that a completely different absorber dominates the T4 spectrum, which was observed one year apart from the others. Although according to the T4 scan plot, a similar high-ionization and fast region exists with a lower significance than that of the best fit, that solution is weakly detected ($\Delta\chi^2/\mathrm{d.o.f.}=13/4$) after including a primary absorber and an emitter (see Sec.\ref{subsubsec:emission}). Therefore, we do not consider that secondary absorption component in the following as the constraints on its parameters are too loose for meaningful discussions.

The flux-resolved spectroscopy is likely to smear/broaden the variable line features, which may lead to degenerate solutions. To reduce the influence of this effect, we fix the line width of \texttt{xabs\_xs} at 9000\,km/s, the average value of the time-resolved results, although the trend discovered below remains the same with a free line width. Among the flux-resolved spectra, we find that a faster, more ionized, and Compton-thicker plasma tends to appear in a brighter state. The corresponding contribution of the UFO to the modeling ($\sim1.2$\,keV) is shown in Fig.\ref{fig:spectrum}. 

\subsubsection{Emission}\label{subsubsec:emission}
The same systematic scan is applied to the \texttt{pion\_xs} model over the continuum model to study the photoionization emission component. The only difference is the searched velocity grid, ranging from $0.1$ to $-0.1$, as we do not find strongly shifted emission lines in the bottom panel of Fig.\ref{fig:ratio-gausian-2018}. The scan result of the 2018 spectrum is shown in the right panel of Fig.\ref{fig:scan} with a fixed line width of 1500\,km/s. It reveals the highly significant detection ($\Delta\chi^2=69$) of a blueshifted ($z_\mathrm{LOS}=-0.011$) photoionized emitter with a modest ionization state ($\log\xi=2.5$). The statistical improvement shown in Tab.\ref{tab:fits} is smaller ($\Delta\chi^2=48$) than that of the scan because some residuals fitted by \texttt{pion\_xs} in the scan over the continuum model have been explained by the model including a \texttt{xabs\_xs}. The primary emitter is not as highly ionized as the absorption component (perhaps related to UFO in T4), since, differently from the absorption component, the photoionization emission is expected to originate from gas located at a wide range of distances from the ionizing source, possibly from large distances. The potential secondary solution ($\log\xi>3.5$) is discussed in Sec.\ref{sec:discussion}.

We do not perform the scan over the time-/flux-resolved spectra as the velocity of the emission component is generally not as variable as the absorption \citep[e.g.][]{2001Kaspi,2016Reeves,2021Kosec}. The best-fit parameters and the contributions of \texttt{pion\_xs} to modeling are shown in Tab.\ref{tab:fits} and Fig.\ref{fig:spectrum} separately. Each solution, except T1, has at least $3.5\sigma$ (i.e. $\Delta\chi^2=20$) detection significance. The unconstrained line width in T1 and T4 is fixed at 1500\,km/s. In general, the line width of the emission component is narrower than that of the absorption, consistent with the expectation of less variable velocity and a larger distance. The column density, ionization state and velocity of the emitter are stable within their uncertainties among the time-resolved spectra, while those parameters are tentatively correlated with the source luminosity in the flux-resolved results. The velocities are all blueshifted at least more than $2700$\,km/s. Apart from F1, the \texttt{pion\_xs} model mainly explains the O \textsc{viii} and Fe/Ne lines around 1\,keV, while it models the O \textsc{vii} and O \textsc{viii} lines in F1 spectra (see Fig.\ref{fig:spectrum}). 

\section{Discussion}\label{sec:discussion}
By analyzing the RGS data of a large \xmm\ campaign on Mrk 1044 in 2018 and 2019, we find a highly-ionized UFO and a blueshifted photoionized emitter in the spectra. The UFO detection confirms the existence of the UFO1 reported in \citet{2021Krongold} from the 2013 \xmm\ observation, sharing a similar ionization state and velocity, although the UFO in their paper was associated with a fixed narrow profile ($\sigma_v=10\,$km/s). The reported multi-phase outflow is also marginally supported by the UFO detected in T4, which has similar parameters to their UFO2 component, although we do not find other cold UFOs like their UFO3/4 phases in Mrk 1044. The emitter shows a much lower ionization state and column density than UFO, implying the average photoionization emission component originates from a different gas with respect to absorption, while the UFO in T4 perhaps is related to the emitter due to their similar column density, ionization state, and turbulent velocity. 

\begin{figure}
	\includegraphics[width=0.5\textwidth, trim={0 150 0 0}]{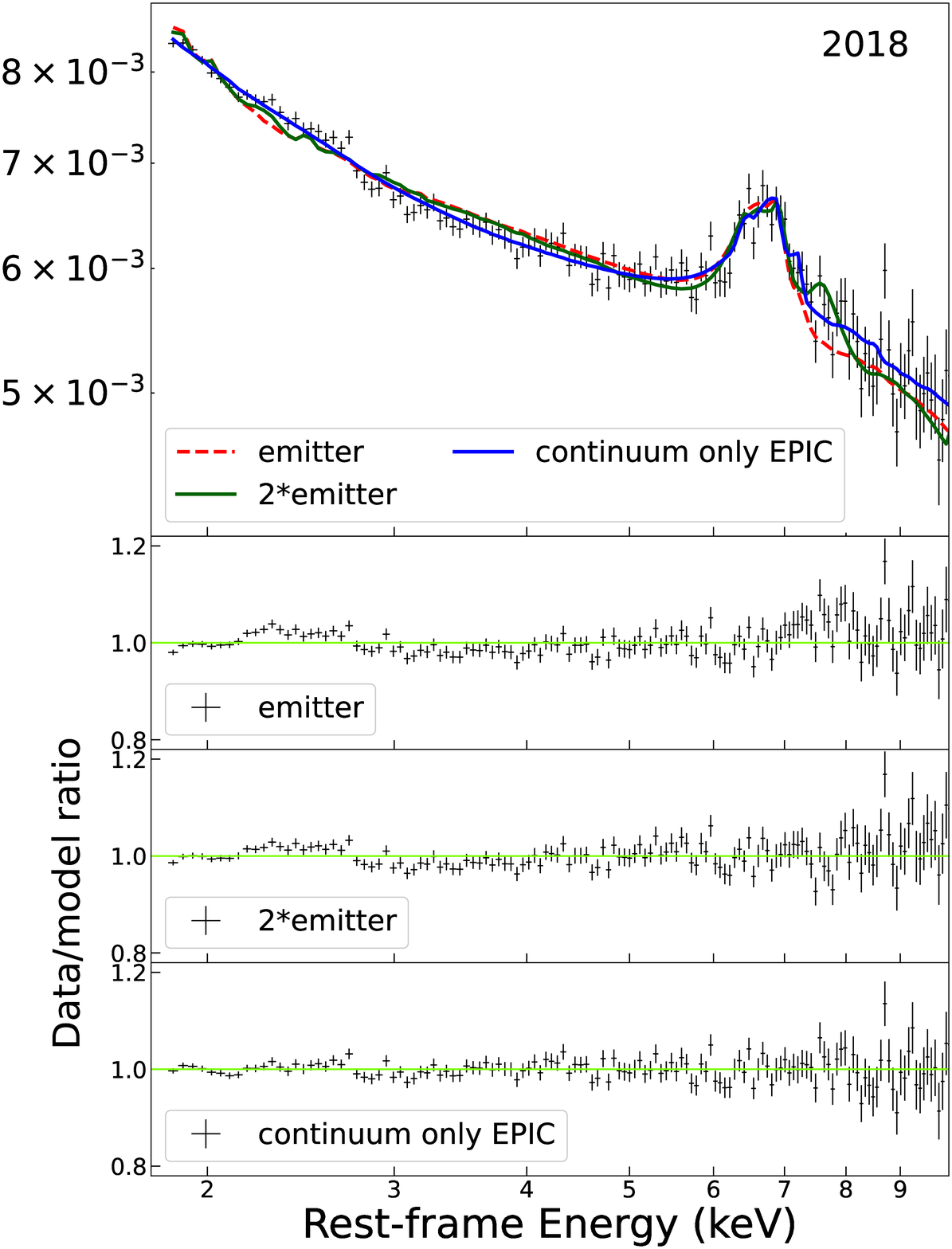}
    \caption{The stacked 2018 EPIC spectra of Mrk 1044. The fits of the continuum plus one or two emitter(s) to the RGS+EPIC data are shown in \textit{red} dashed and \textit{green} solid lines, respectively. The fit of the continuum model to only EPIC data ($1.77\mbox{--}10$\,keV for consistency with other fits) is shown in the \textit{blue} line. The corresponding data/model ratios are shown in the following panels.}
    \label{fig:EM_2018}
\end{figure}

In the scan of the photoionization emission model (see the right panel of Fig.\ref{fig:scan}), we also discover a potential secondary emitter, which is trying to complement the blue wing of the Fe K emission. The best fit ($\Delta\chi^2=37$) of the secondary emitter requires an ultra-fast ($z_\mathrm{LOS}=-0.12^{+0.02}_{-0.02}$) and highly ionized ($\log\xi=3.7^{+0.1}_{-0.2}$) plasma with a column density of $N_\mathrm{H}=3.6^{+1.8}_{-1.5}\times10^{21}\,\mathrm{cm}^{-2}$ and an unconstrained turbulent velocity fixed at $\sigma_v=9000$\,km/s. This emitter shares common properties with the absorber, indicating the same origin of the absorption. We plot the stacked 2018 EPIC spectra and the best-fit model including two emitters in the top panel of Fig.\ref{fig:EM_2018}, compared with the continuum plus one emitter. The corresponding data-to-model ratios are shown in the second and third panels. However, we are concerned about the requirement of this secondary emitter as the not well-explained Fe K profile might result from the imbalance between the statistics of the RGS and EPIC data, where the grating data have more bins (a factor of 4) than CCD data and the model is mainly adjusted to fitting soft X-ray residuals. To test this possibility, we fit the continuum model to only EPIC data and show it in Fig.\ref{fig:EM_2018} as well as the corresponding ratio. Compared with the results in Tab.\ref{tab:fits}, the continuum model only requires a harder spectral slope ($\Gamma=2.16^{+0.03}_{-0.02}$) and explains the blue wing of Fe emission well, while the other parameters remain unchanged within uncertainties. In terms of fitting the EPIC data, this continuum model is much better ($\Delta\chi^2=86$) than the continuum plus two emitters fitted to the RGS+EPIC data. It suggests that the additional emission component is spurious, although we cannot exclude the possibility of an intervening outflow contributing to a part of the Fe K profile. Therefore, we tend not to discuss the evolution of the secondary emitter in the following. 

\subsection{Evolution of the wind components}\label{subsec:relation}
\begin{figure*}
	\includegraphics[width=0.45\textwidth, trim={0 30 0 0}]{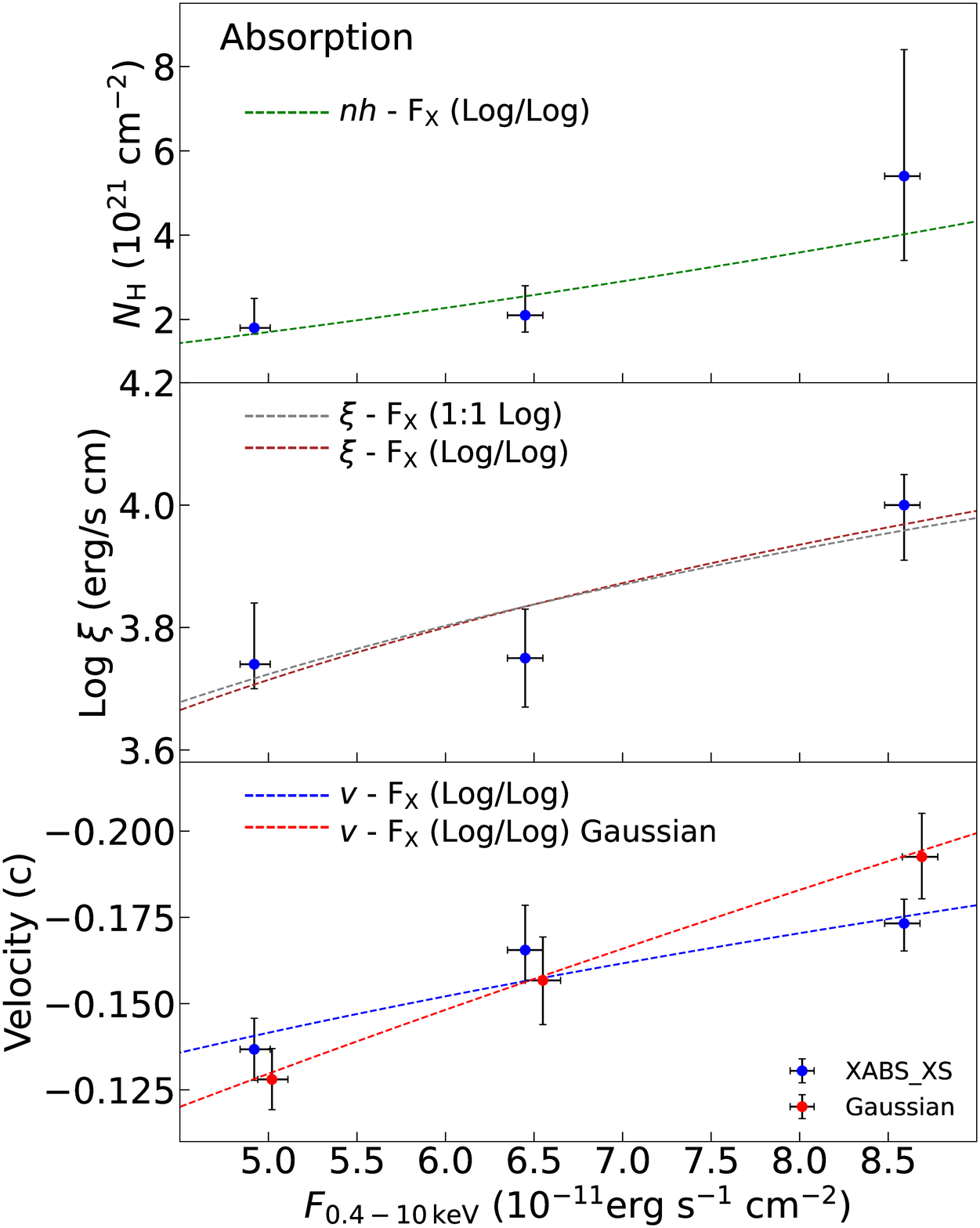}
	\includegraphics[width=0.45\textwidth, trim={0 30 0 0}]{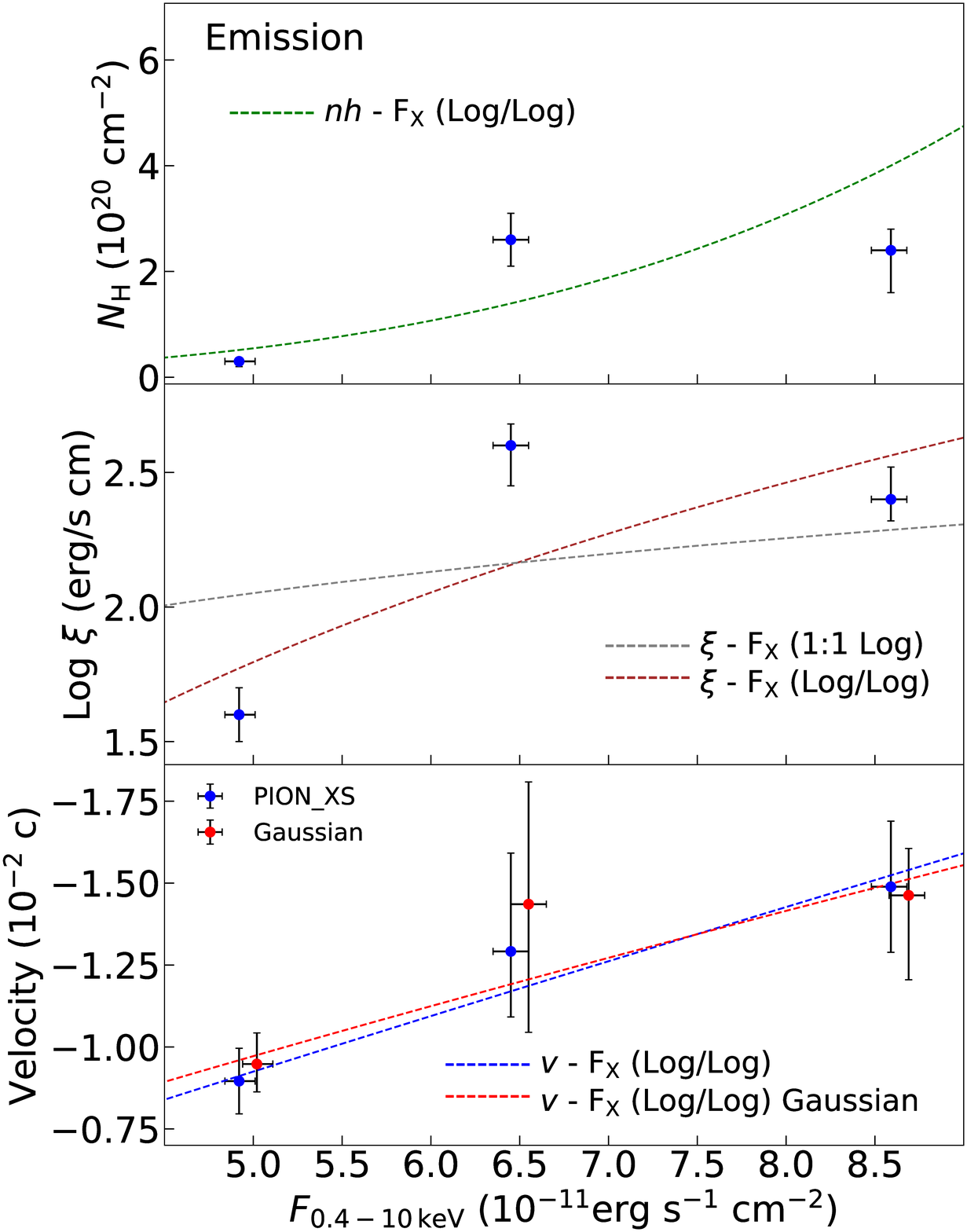}
    \caption{The column density ({\it top}), ionization parameter ({\it middle}) and velocity ({\it bottom}) of the photoionized absorbing ({\it left}) and emitting ({\it right}) plasmas versus the unabsorbed X-ray flux for the flux-resolved spectra. The blueshift of the main absorption/emission feature (i.e. Ne \textsc{x} / O \textsc{viii}) measured by \texttt{Gaussian} is also included, where the corresponding flux is manually shifted for clarity. The linear function fits with (1:1 Log) and without (Log/Log) a slope fixed at unity are performed in a logarithmic space. See details in Sec.\ref{subsec:relation}.}
    \label{fig:evolution}
\end{figure*}

In Sec.\ref{subsec:scan}, we have measured the properties of the absorption and emission components at different flux levels. To further investigate the relations between wind properties and source luminosity, we plot their column density, ionization parameter, and velocity versus the calculated fluxes in Fig.\ref{fig:evolution}. The blueshift of the absorption and emission feature, measured by the \texttt{Gaussian} model, are included as well. The absorption line is assumed to come from Ne \textsc{x} ($\mathrm{E}_\mathrm{rest}=1.022$\,keV), while the emission line is O \textsc{viii} ($\mathrm{E}_\mathrm{rest}=0.6535$\,keV). We fit these parameters with a linear function in a logarithmic space. The same fit with a slope fixed at unity is also performed on the ionization parameter to show the expected behavior in photoionization equilibrium, according to the definition of the ionization parameter ($\xi\equiv L_\mathrm{ion}/n_\mathrm{H}R^2\propto F_\mathrm{ion}$, where $R$ is the distance from the ionizing source to the plasma and $n_\mathrm{H}$ is the hydrogen volume density). All of the fits provide positive correlations between wind properties and the source luminosity.
\subsubsection{Absorbing gas}\label{subsubsec:absorption-evolution}
For the absorption component, the Pearson correlation coefficients of the best-fit values of ($N_\mathrm{H}, F$), ($\log\xi, F$), ($v, F$), ($v_\mathrm{gaus}, F$) points considering their uncertainties are 0.76, 0.86, -0.86 and -0.96 respectively \citep{2014Curran}, suggesting a moderate correlation between ($N_\mathrm{H}, F$) and strong correlation among the others. The Log/Log fits give:
\begin{equation}\label{eq:abs-nh}
   \log\frac{N_\mathrm{H}}{10^{21}\,\mathrm{cm}^{-2}}= (-0.9\pm0.8)+(1.59\pm0.98)\log(\frac{F_\mathrm{0.4-10}}{10^{-11}}),
\end{equation}
\begin{equation}\label{eq:abs-ionization}
    \log\frac{\xi}{\mathrm{erg\,cm/s}}= (2.96\pm0.33)+(1.08\pm0.41)\log(\frac{F_\mathrm{0.4-10}}{10^{-11}}), 
\end{equation}
\begin{equation}\label{eq:abs-v}
 \log\frac{|v|}{\mathrm{c}}= (-1.12\pm0.14)+(0.39\pm0.16)\log(\frac{F_\mathrm{0.4-10}}{10^{-11}}),
\end{equation}
\begin{equation}\label{eq:abs-v_gaus}
  \log\frac{|v_\mathrm{gaus}|}{\mathrm{c}}= (-1.40\pm0.16)+(0.73\pm0.19)\log(\frac{F_\mathrm{0.4-10}}{10^{-11}}).
\end{equation}
The best-fit value of the slope in Eq.\ref{eq:abs-ionization} is consistent with one, which is the expected value through the definition, in spite of the large uncertainty, indicating that the absorbing gas responds to the variability of the source radiation instantaneously, which implies a high volume density. 
The weakly increasing trend of the column density is opposite to the relation shown in IRAS 13224-3809 \citep[see Fig.7 in][]{2018Pinto}, where the column density slightly decreases along with the increasing ionization parameter ($\log\xi$ up to 6) and luminosity. It means the UFO in Mrk 1044 may have not been over-ionized, suggested by the modest ionization state ($\log\xi\sim3.7-4.0$), and still require a larger column density to visualize the absorption features at higher ionization state \citep[e.g. see Fig. 10 in][]{2020Pinto}. 

The positive correlation between the velocity and the X-ray flux suggests that the wind is radiatively driven, which is also observed in other high-accretion systems, IRAS 13224-3809 \citep{2018Pinto} and PDS 456 \citep{2017Matzeu}. According to Eq.4 in \citet{2017Matzeu}, the net radiative-driven (i.e. radiative minus gravitational force) outflow should have a dependence between the velocity, the luminosity $L_\mathrm{ion}$ and the launching radius $R_\mathrm{w}$, 
\begin{equation}\label{eq:L-v}
v/c\propto k_\mathrm{0.4-10}^{1/2}L_\mathrm{0.4-10}^{1/2}R_\mathrm{w}^{-1/2},
\end{equation}
where $k_\mathrm{0.4-10}=L_\mathrm{bol}/L_\mathrm{0.4-10}$ is the bolometric correction factor. The relation observed in Mrk 1044 (from \texttt{xabs\_xs} instead of the phenomenological model) is consistent with the power index (0.5) in Eq.\ref{eq:L-v} within uncertainties, at variance with the results derived from IRAS 13224-3809 ($0.05\pm0.02$) and PDS 456 ($0.22\pm0.04$). 

\subsubsection{Emitting gas}\label{subsubsec:emission-evolution}
For the emission component, the Pearson correlation coefficients of the best-fit values of ($N_\mathrm{H}, F$), ($\log\xi, F$), ($v, F$), ($v_\mathrm{gaus}, F$) points considering their uncertainties are 0.73, 0.68, -0.83, and -0.69, respectively, suggesting moderate correlations, except for a strong correlation between ($v, F$). The fits provide:
\begin{equation}\label{eq:em-nh}
    \log\frac{N_\mathrm{H}}{10^{20}\,\mathrm{cm}^{-2}}= (-2.8\pm2.3)+(3.68\pm2.79)\log(\frac{F_\mathrm{0.4-10}}{10^{-11}}),
\end{equation}
\begin{equation}\label{eq:em-ionization}
  \log\frac{\xi}{\mathrm{erg\,cm/s}}= (-0.49\pm2.16)+(3.27\pm2.64)\log(\frac{F_\mathrm{0.4-10}}{10^{-11}}), 
\end{equation}
\begin{equation}\label{eq:em-v}
    \log\frac{|v|}{0.01\mathrm{c}}= (-0.68\pm0.14)+(0.92\pm0.17)\log(\frac{F_\mathrm{0.4-10}}{10^{-11}}),
\end{equation}
\begin{equation}\label{eq:em-v_gaus}
   \log\frac{|v_\mathrm{gaus}|}{0.01\mathrm{c}}= (-0.57\pm0.17)+(0.80\pm0.22)\log(\frac{F_\mathrm{0.4-10}}{10^{-11}}).
\end{equation}
However, the extremely large uncertainties of the fitted parameters, except for the velocity-related fits, preclude any authentic conclusions on the emitting gas. 
It is also noted that the timescale of the segments of the flux-resolved spectra, around 3\,ks, provides a maximal distance of $\sim220\,R_\mathrm{g}$ for the causally linked correlations, unless the emitting gas is mainly located along with our LOS, which is rather unlikely but still possible (given its blueshift). However, the large range of locations of the emission, suggested by the moderate ionization state, a low column density, and a narrow line width, will result in a low coherence between the plasma and the source \citep{2022Juranova}, impeding the discovery of correlations. The observed variation of the velocity is, therefore, only probably contributed by a portion of the emitting gas near the central region.

\subsection{Outflow properties}\label{subsec:kinetic}
It is expected that outflows carry out sufficient power to quench or trigger star formation in their hosts and affect the evolution of galaxies \citep[e.g.][]{2005DiMatteo,2010Hopkins,2017Maiolino,2022Chen}. According to simulations, the deposition of the kinetic energy larger than $0.5\%$ of the Eddington luminosity into the ISM is sufficient to produce considerable feedback on the host galaxy. The kinetic power of UFO can be expressed as:
\begin{equation}\label{eq:UFO-kinetic}
     L_\mathrm{UFO}=\frac{1}{2}\dot{M}_\mathrm{out}v^2_\mathrm{UFO}=\frac{1}{2}\Omega R^2\rho v_\mathrm{UFO}^3 C_\mathrm{V},
\end{equation}
where $\dot{M}_\mathrm{out}=\Omega R^2\rho v_\mathrm{UFO}C$ is the mass outflow rate; $\Omega$ the opening angle; $R$ the distance between the ionizing source and UFO; $\rho$ the outflow mass density; $C_\mathrm{V}$ the volume filling factor. The mass density is defined as $\rho=n_\mathrm{H}m_\mathrm{p}\mu$, where $n_\mathrm{H}$ is the hydrogen number density; $m_\mathrm{p}$ the proton mass and $\mu=1.2$ the mean atomic mass assuming solar abundances. $n_\mathrm{H}R^2$ can be replaced by obtained parameters $L_\mathrm{ion}/\xi$, according to the definition of the ionization parameter ($\xi\equiv L_\mathrm{ion}/n_\mathrm{H}R^2$). We estimate the ionizing luminosity ($1\mbox{--}1000$\,Rydberg) from the SED presented in Fig.\ref{fig:SED} at $L_\mathrm{ion}\sim3.9\times10^{43}\,\mathrm{erg/s}$ and find,
\begin{equation}\label{UFO-kinetic-2}
L_\mathrm{UFO}=0.5v_\mathrm{UFO}^3m_\mathrm{p}\mu L_\mathrm{ion}\Omega C/\xi\sim5.89\times10^{44}\Omega C_\mathrm{V}\,\mathrm{erg/s}
\end{equation}
by inputting the results of the UFO obtained in the 2018 spectrum. Here we adopt a conservative value of the opening angle $\Omega=0.3$ from the GRMHD simulations of radiative-driven outflows in high-accretion systems \citep{2013Takeuchi}. The filling factor $C_\mathrm{V}=7\times10^{-3}$ is derived from Eq.23 in \citet{2018Kobayashi} assuming that the outflow mass rate is comparable with the accretion mass rate and the accretion efficiency is $\eta=0.1$. The conservative value of the UFO kinetic energy is thus $L_\mathrm{UFO}\sim1.54\times10^{43}\,\mathrm{erg/s}\sim4.4\%L_\mathrm{Edd}$, surpassing the theoretical criterion, suggesting that the UFO in Mrk 1044 is very likely to influence the evolution of the host galaxy.

\begin{figure}
	\includegraphics[width=0.45\textwidth, trim={0 20 0 0}]{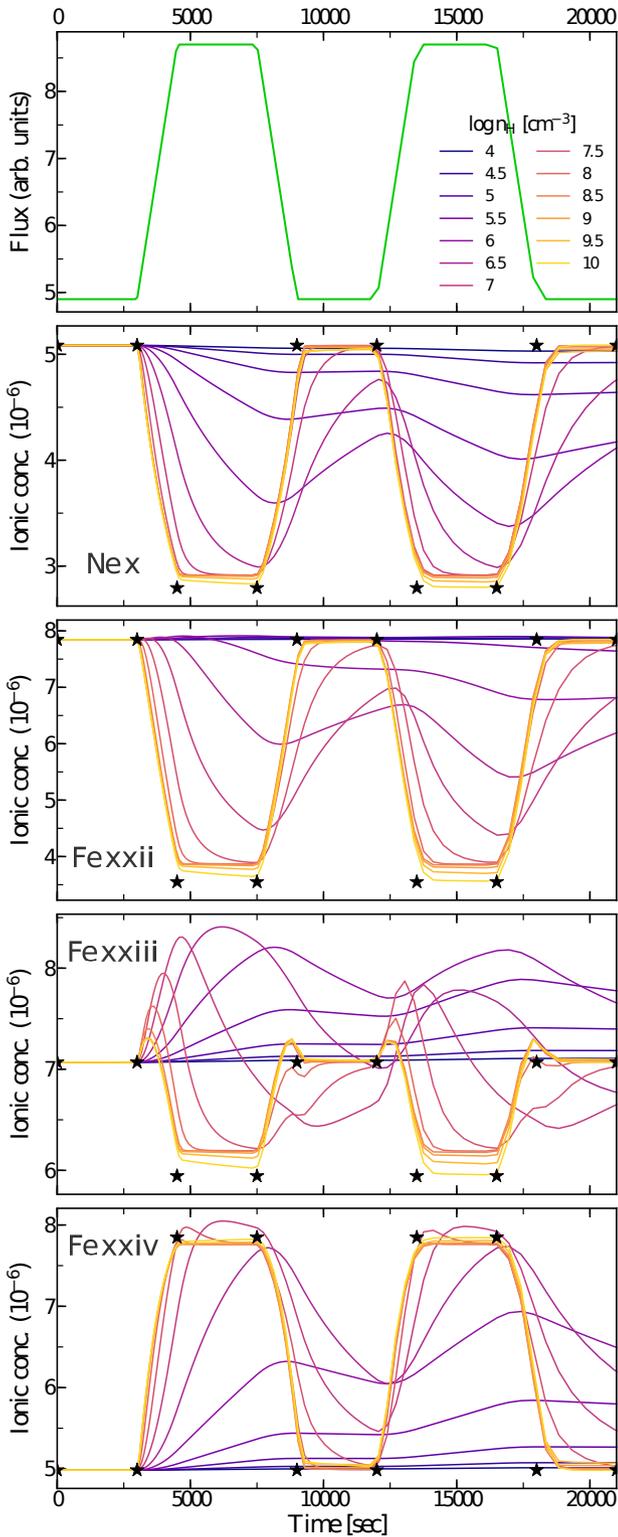}
    \caption{\textit{Top} panel: The input light curve that we expect in Mrk 1044 (an approximation) for the \texttt{tpho} model. The low state corresponds to the luminosity of F1 and the high state to the luminosity of F3. The duration of the low state, the middle, and the high state are 3\,ks, 1.5\,ks, and 3\,ks respectively, which are the average timescale of the segments of the flux-resolved spectra. \textit{Middle} and \textit{Botton} panels: The time-dependent evolution of the concentration relative to hydrogen of Ne \textsc{x} and Fe \textsc{xxii-xxiv} for different gas densities and are compared with the ionic concentrations for a plasma in photoionization equilibrium (black stars).} 
    \label{fig:density}
\end{figure}

Based on the hypothesis that the UFO velocity is at least larger than its escape velocity, we can estimate the lower limit of the outflow location, $R\geq2GM_\mathrm{BH}/v_\mathrm{UFO}^2\geq98R_\mathrm{g}$. It provides an upper limit of the outflow density, $n_\mathrm{H}=L_\mathrm{ion}/\xi R^2<4.5\times10^{12}\,\mathrm{cm}^{-3}$. On the other hand, 
by using the time-dependent photoionization model \texttt{tpho} \citep{2022Rogantini}, we simulate the response of plasma with different densities to the source variability to estimate the lower limit of the plasma density and further the upper limit of the outflow location. The duration of the low, middle, and high states of the source are 3\,ks, 1.5\,ks, 3\,ks respectively, provided by the timescale of segments of flux-resolved spectra, shown in the top panel of Fig.\ref{fig:density}. The time-dependent evolution of the ionic concentration of predominant absorption lines, i.e., Ne \textsc{x} and Fe \textsc{xxii-xxiv}, are shown in the lower four panels of Fig.\ref{fig:density}. Gases with a density above $10^{9}\,\mathrm{cm}^{-3}$ respond quickly to the luminosity. If we assume the UFO in Mrk 1044 responds to the source instantaneously, suggested by Eq.\ref{eq:abs-ionization}, the lower limit of the density is $10^{9}\,\mathrm{cm}^{-3}$. The recombination timescale of the plasma at $\log\xi=3.7$ can be evaluated through the \texttt{rec\_time} code in SPEX. For example, the recombination time of the Fe \textsc{xxiv} line, the predominant line in UFO, is calculated $t_\mathrm{rec}<19$\,s, consistent with our assumption.
The volume density of the UFO is thus estimated between $n_\mathrm{H}=10^9\mbox{--}4.5\times10^{12}\,\mathrm{cm}^{-3}$ and the corresponding location is $R=\sqrt{L_\mathrm{ion}/n_\mathrm{H}\xi}=98\mbox{--}6600\,R_\mathrm{g}$.

For the emitting gas, we have to adopt another method to derive the upper limit of its location, since its response to the source is unconstrained. Through the assumption that the plasma thickness is smaller than its distance to the source, $\Delta R=N_\mathrm{H}/C_\mathrm{V}n_\mathrm{H}\leq R$, the upper limit of the location can be estimated, $R\leq C_\mathrm{V}L_\mathrm{ion}/\xi N_\mathrm{H}\leq7.8\times10^{6}\,R_\mathrm{g}$. The lower limit could be calculated by the same method for absorbing gas, based on the same assumption of $v_\mathrm{LOS}\geq v_\mathrm{esc}$, at $>1.2\times10^{4}\,R_\mathrm{g}$.
However, since the emission comes from a wide range of distances, the observed velocity is an averaged value and may not be representative of the escape velocity of the emitting gas close to the center. If we assume the emitting gas shares the same origin with the UFO detected in T4, the lower limit of the location can be estimated at $>320\,R_\mathrm{g}$, close to the maximal distance of the causal connection. Our constraints on the location of the emission components are rather loose and span the whole range of distances, $3\times10^{2}\mbox{--}7.8\times10^{6}\,R_\mathrm{g}$, from the accretion disk to the interface between the outer disk and the broad line region (BLR), which scales with the black hole mass or Eddington ratio. The range of the corresponding density is $1.2\times10^4\mbox{--}7\times10^{11}\,\mathrm{cm}^{-3}$. 

\subsection{Comparison with other AGN}\label{subsec:comparison}
By comparison with UFOs discovered in other AGN, the ionization state ($\log\xi\sim3.7$) and the velocity ($v\sim0.15c$) of the UFO in Mrk 1044 are typical \citep[$\log\xi\sim3\mbox{--}6$, $v\sim0.08\mbox{--}0.3c$, e.g.][]{2015Nardini,2020Kosec,2021Parker,2021Xu,2022Matzeu}. The column density ($N_\mathrm{H}\sim2.3\times10^{21}\,\mathrm{cm}^{-2}$) is not as thick as typical UFOs discovered from Fe K absorption feature ($\log (N_\mathrm{H}/\mathrm{cm}^{-2})\sim22\mbox{--}24$). However, the low column density is common in UFOs detected in the soft X-ray band \citep[e.g.][]{2015Longinotti,2016Pounds,2022Xu}. Alternatively, another potential explanation is the relatively low inclination angle of Mrk 1044 ($i\sim34^\circ$) that we are therefore viewing a narrower wind region.

The correlation between the velocity of the UFO and the source luminosity is consistent with the phenomenon observed in PDS 456 \citep{2017Matzeu} and IRAS 13224-3809 \citep{2018Pinto}, while different from 1H 0707-495 \citep{2021Xu}. The reason might come from their different Eddington ratios, as the former three (PDS 456, $\lambda_\mathrm{Edd}\sim0.77$, \citealt{2015Nardini}; IRAS 13224-3809, $\lambda_\mathrm{Edd}=1\mbox{--}3$, \citealt{2019Alston}; Mrk 1044, $\lambda_\mathrm{Edd}\sim0.4$) are not so highly accreting as 1H 0707-495 ($\lambda_\mathrm{Edd}>0.7$, \citealt{2021Xu}, or $\lambda_\mathrm{Edd}=140\mbox{--}260$, \citealt{2016Done}) of which the structure of the accretion flow does not significantly deviate from the standard thin disk model \citep{1973Shakura}.
\begin{figure}
	\includegraphics[width=0.45\textwidth, trim={0 30 0 0}]{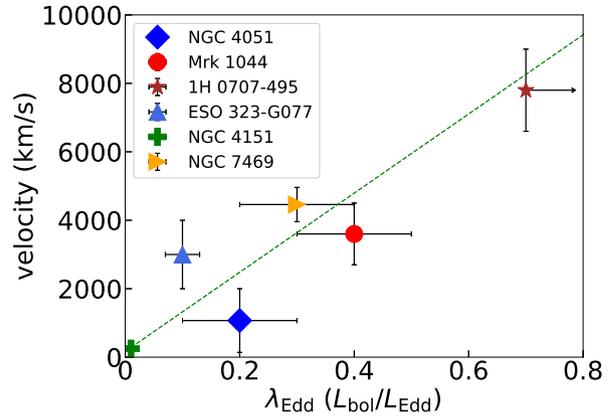}
    \caption{The velocity of the emission lines versus the estimated Eddington ratio of Type 1 AGN. See more details in Sec.\ref{subsec:comparison}.}
    \label{fig:mdot_v}
\end{figure}

Blue-shifted emission lines are rarely observed in AGN. To our knowledge, among Type 1 AGN, only four sources (1H 0707-495, \citealt{2021Xu}; ESO 323-G077, \citealt{2008Jim}; NGC 4151, \citealt{2007Armentrout}; NGC 7469, \citealt{2020Grafton}) reveal blueshifted emission lines, as well as the partially absorbed emission lines in NGC 4051 \citep{2011Pounds}. 
Given the fact that blueshifted emission lines were also found in some Ultra-luminous X-ray (ULX) sources \citep[e.g., NGC 55 ULX and NGC 247 X-1,][]{2017Pinto,2021Pinto,2021Kosec}, we propose that blueshifted emission lines are related to high accretion rates and plot the blueshift of emission lines versus the Eddington ratios in Fig.\ref{fig:mdot_v} \citep{2008Jim,2017Edelson,2018Mehdipour,2021Xu,2021Yuan}. The Eddington ratio of 1H 0707-495 is assumed at its lower limit, 0.7, as we cannot constrain the upper limit and only know it is a super-Eddington AGN. The Pearson coefficient is 0.87, suggesting they are highly correlated. The linear fit gives:
\begin{equation}\label{mdot-blueshift}
v_\mathrm{EM}(\mathrm{km/s})=(11560\pm1756)\lambda_\mathrm{Edd}+(170\pm240).
\end{equation}
Although the fit seems to strongly support our hypothesis, due to the small size of the sample and the uncertainty on the $\lambda_\mathrm{Edd}$, we are unable to confirm that correlation. The reason why there are no blueshifted emission lines in other high-Eddington AGN is probably a too-strong continuum, washing out the lines, or the small viewing angle close to face-on. The validation of this correlation requires a systematic analysis like what we have done in this paper on a large sample of AGN at different accretion rates.

\subsection{Future Mission}\label{subsec:future}
Future missions with unprecedented spectral resolution and effective area will provide incredible constraints on the nature of UFOs. Their large effective area will collect sufficient photons for spectroscopy within a short timescale, thus avoiding the risk of spectral broadening due to the flux-resolved spectroscopy. We will be able to trace the variability of UFO properties at different flux levels and put tighter constraints on the region of outflows through the variability ($\Delta R=c\Delta t$).

\begin{figure}
	\includegraphics[width=0.5\textwidth, trim={0 80 0 0}]{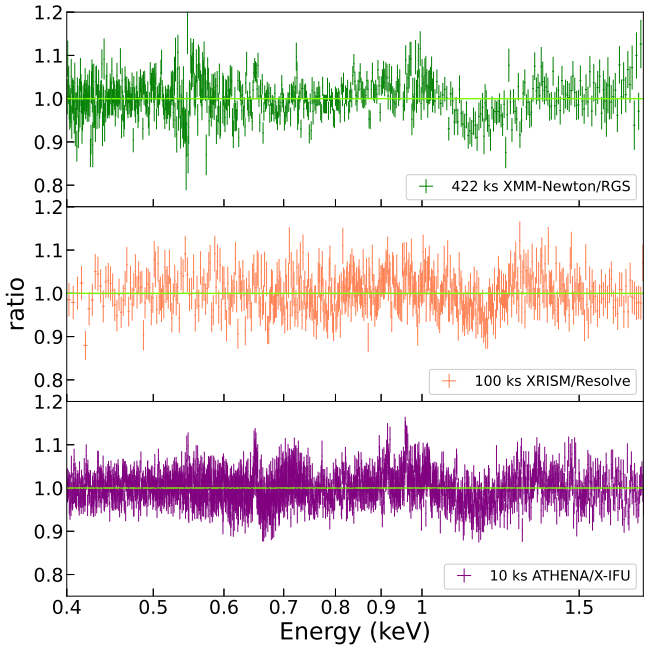}
    \caption{The data/model ratio for the XRISM/Resolve (\textit{middle}, 100\,ks) and ATHENA/X-IFU (\textit{bottom}, 10\,ks) spectrum, simulated by the best-fit model obtained in Sec.\ref{subsec:scan}, with the respect to the baseline continuum model. The ratio for the stacked 422\,ks \xmm\ 2018 spectrum is presented in the \textit{top} panel for comparison.}
    \label{fig:future}
\end{figure}

We, therefore, simulate spectra for the X-Ray Imaging and Spectroscopy Mission \citep[\xrism,][]{2018Tashiro} and the Advanced Telescope for High-Energy Astrophysics \citep[\athena,][]{2013Nandra} based on the best-fit model obtained in Sec.\ref{subsec:scan}. The data/model ratios with respect to the continuum model are shown in the middle and bottom panel of Fig.\ref{fig:future} and that of the stacked \xmm\ 2018 spectrum is presented in the top panel for comparison. Compared with the detection significance of an outflow and an emitter in \xmm\ spectrum ($\Delta\chi^2=203$), we find \xrism\ provides a comparable statistical improvement ($\Delta\chi^2=205$) with a quarter of \xmm\ exposure time (100\,ks), while \athena\ can reach a much stronger detection ($\Delta\chi^2=1020$) within over one order of magnitude fewer exposure time (10\,ks).


Furthermore, the huge spectral resolution is likely to resolve the line profile, which conceals the information about the launching mechanism. According to \citet{2022Fukumura}, the outflows driven by radiative lines have an asymmetric line shape of an extended red wing, while those driven by magnetic fields have a blue-extended wing. Such a difference is able to be distinguished with high-resolution missions. Therefore, given these two benefits, it is promising to deepen our understanding of AGN outflows and identify the UFO launching mechanism with future missions. Moreover, an emulation method recently developed by \citet{2022Matzeub} can be applied to efficiently model the spectral data from future missions, which have the capability to collect photons orders of magnitude larger than that of current facilities.

\section{Conclusions}\label{sec:conclusion}
In this work, through the time- and flux-resolved X-ray spectroscopy on four \xmm\ observations of Mrk 1044, we investigate the dependence of the wind properties on the source luminosity. We find that the absorbing gas quickly responds to the source variability, suggesting a high-density plasma ($n_\mathrm{H}\sim10^{9}\mbox{--}4.5\times10^{12}\,\mathrm{cm}^{-3}$). Furthermore, the UFO velocity is correlated with the X-ray luminosity, suggesting that the UFO in Mrk 1044 is accelerated by the radiation field. The emitting gas is located at a large range of distances from the SMBH and shows a blueshift of $2700\mbox{--}4500$\,km/s. By comparing with the discovered blueshifted emission lines in other AGN, we propose that the blueshift of emission lines is probably correlated with the source accretion rate, which can be verified with a large sample study. Our simulations demonstrate that the nature of AGN winds will be promisingly unveiled by future missions due to their large effective area and high spectral resolution.
\section*{Acknowledgements}
D.R. is supported by NASA through the Smithsonian Astrophysical Observatory (SAO) contract SV3-73016 to MIT for Support of the Chandra X-Ray Center (CXC) and Science Instruments. S.B. acknowledges financial support from the Italian Space Agency under the grant ASI-INAF 2017-14-H.O. E.K. acknowledges XRISM Participating Scientist Program for support under NASA grant 80NSSC20K0733. C.J. acknowledges the National Natural Science Foundation of China through grant 11873054, and the support by the Strategic Pioneer Program on Space Science, Chinese Academy of Sciences through grant XDA15052100.


\section*{Data Availability}
The \xmm\ data in this article are available in ESA’s XMM-Newton Science Archive (https://www.cosmos.esa.int/web/xmm-newton/xsa).


\bibliographystyle{mnras}
\bibliography{ref} 







\bsp	
\label{lastpage}
\end{document}